\documentclass[10pt,prd,preprintnumbers,floatfix,aps,nofootinbib,noshowpacs,twocolumn,superscriptaddress]{revtex4-1}

\usepackage{epsfig}
\usepackage{url}
\usepackage{hyperref}

\usepackage[normalem]{ulem} 

\usepackage{latexsym}
\usepackage{epsfig}
\usepackage{amsmath}
\usepackage{amssymb}
\usepackage{wasysym}
\usepackage{graphicx}
\usepackage{verbatim}
\usepackage{enumerate,mdwlist}%lists
\usepackage[titletoc]{appendix}
\usepackage{amsfonts}
\usepackage{tikz} %diagrams

\usepackage[normalem]{ulem}%strikethrough text

\newcommand{\ba}{\begin{eqnarray}}
\newcommand{\ea}{\end{eqnarray}}
\newcommand{\be}{\begin{equation}}
\newcommand{\ee}{\end{equation}}

\definecolor{grey}{rgb}{0.4,0.4,0.4}
\definecolor{dullmagenta}{rgb}{0.4,0,0.4}
\definecolor{darkblue}{rgb}{0,0,0.4}
\definecolor{midblue}{rgb}{0,0,0.5}
\definecolor{midred}{rgb}{0.5,0,0}
\definecolor{orange}{rgb}{1,0.5,0}
\definecolor{lightbrown}{rgb}{0.75,0.5,0.25}
\definecolor{tan}{cmyk}{0.14,0.42,0.56,0}
\definecolor{djunglegreen}{cmyk}{0.99,0,0.52,0}
\definecolor{lightgreen}{rgb}{0,1,0}
\definecolor{olivegreen}{cmyk}{0.64,0,0.95,0.40}
\definecolor{midgreen}{rgb}{0.0,0.675,0.0}
\definecolor{darkgreen}{rgb}{0,0.5,0}

\usepackage[all]{xy} %array diagrams
\usepackage{amsfonts}

\newcommand{\papertitle}{
Limits on stellar-mass compact objects as dark matter \\
from gravitational lensing of type Ia supernovae
}
\begin{document} 

\title{\papertitle} 

\author{Miguel Zumalac\'arregui}
\email{miguelzuma@berkeley.edu}
\affiliation{Berkeley Center for Cosmological Physics, LBNL and University of California at Berkeley, \\
Berkeley, California 94720, USA}
\affiliation{Institut de Physique Th\' eorique, Universit\'e  Paris Saclay 
CEA, CNRS, 91191 Gif-sur-Yvette, France}
\affiliation{Nordita, KTH Royal Institute of Technology and Stockholm University, Roslagstullsbacken 23, SE-106 91 Stockholm, Sweden}

\author{Uro\v{s} Seljak}
\email{useljak@berkeley.edu}
\affiliation{Berkeley Center for Cosmological Physics, LBNL and University of California at Berkeley, \\
Berkeley, California 94720, USA}
\affiliation{Physics and Astronomy Department, LBNL, University of California at Berkeley, \\
Berkeley, California 94720, USA}

\begin{abstract}

The nature of dark matter (DM) remains unknown despite very precise knowledge of its abundance in the universe.
An alternative to new elementary particles postulates DM as made of macroscopic compact halo objects (MACHO)
such as black holes formed in the very early universe. 
Stellar-mass primordial black holes (PBHs) are subject to less robust constraints than other mass ranges and might be connected to gravitational-wave signals detected by the Laser Interferometer Gravitational-Wave Observatory (LIGO).
New methods are therefore necessary to constrain the viability of compact objects as a DM candidate.
Here we report bounds on the abundance of compact objects from gravitational lensing of type Ia supernovae (SNe). 
Current SNe datasets constrain compact objects to represent less than 35.2\% (Joint Lightcurve Analisis) and 37.2\% (Union 2.1) of the total matter content in the universe, at 95\% confidence-level.
The results are valid for masses larger than $\sim 0.01 M_\odot$ (solar-masses), 
limited by the size SNe relative to the lens Einstein radius.
We demonstrate the mass range of the constraints by computing magnification probabilities for realistic SNe sizes and different values of the PBH mass.
Our bounds are sensitive to the total abundance of compact objects with $M \gtrsim 0.01 M_\odot$ and complementary to other observational tests.
These results are robust against cosmological parameters, outlier rejection, correlated noise and selection bias.
PBHs and other MACHOs are therefore ruled out as the dominant form of DM for objects associated to LIGO gravitational wave detections.
These bounds constrain early-universe models that predict stellar-mass PBH production and strengthen the case for lighter forms of DM, including new elementary particles.
\end{abstract}

\date{\today}

\keywords{}

\maketitle

\section{Introduction}

%DM basics
A major goal of cosmology is to characterize the physical constituents and laws of the universe. The nature of Dark Matter (DM), the component sourcing the formation of large scale structure (LSS) and contributing 27\% of the energy budget of the universe \cite{Ade:2015xua}, remains highly elusive despite decades of dedicated searches. Standard DM scenarios postulate a new elementary particle, abundantly produced in the early universe and whose interaction with standard model particles is sufficiently suppressed, in agreement with bounds on detection experiments and collider production rates \cite{Bertone:2004pz}. 
Cosmological observations are insensitive to microscopic details of DM as long it behaves as a non-relativistic fluid, or cold dark matter (CDM), on large scales.

%MACHOs and PBH
An alternative to microscopic dark matter scenarios invokes Primordial Black Holes (PBH) formed in the early universe \cite{1967SvA....10..602Z,Carr:1974nx,Meszaros:1974tb,1975Natur.253..251C} or other macroscopic entities, generically known as massive compact halo objects (MACHO).
PBHs behave as non-relativistic matter on large scales, making them cosmologically viable CDM candidates. They are neither detected nor produced in particle physics experiments, but can be probed by a series of small-scale effects that depend on the mass and properties of the objects \cite{Carr:2016drx,Sasaki:2018dmp,Carr:2018rid}, see Fig. \ref{fig:mass_dep_constraints}.

%LIGO band
Interestingly, the less robust constraints on PBHs  $M\sim 10-100 M_\odot$ (solar-mass) coincide with the masses of black holes detected by the Laser Interferometer Gravitational-wave Observatory (LIGO) \cite{Abbott:2016blz,TheLIGOScientific:2016pea}.
This intriguing possibility lead to a revival of PBH models \cite{Bird:2016dcv,Clesse:2016vqa} that could simultaneously satisfy existing bounds, provide the right dark matter abundance and explain the high merger rate and progenitor masses inferred by the first LIGO detections. 
Unfortunately, uncertainties in the small-scale distribution of PBHs remain an obstacle to constrain their abundance on the basis of current gravitational wave observations alone (although see \cite{Nakamura:1997sm,Sasaki:2016jop,Ali-Haimoud:2017rtz,Chen:2018czv,Magee:2018opb,Abbott:2018oah}).

Other methods are needed to reliably test the PBH-DM hypothesis.
%Our job
In this letter we explore the  gravitational lensing predictions of PBHs-DM models for type Ia supernovae (SNe) as standard candles whose luminosity can be calibrated.
Our analysis using current SNe datasets improves considerable on previous bounds \cite{Metcalf:2006ms}.
Details of the analysis are presented in the Supplemental Material.

\begin{figure*}[t!]
\includegraphics[width=0.99\textwidth]{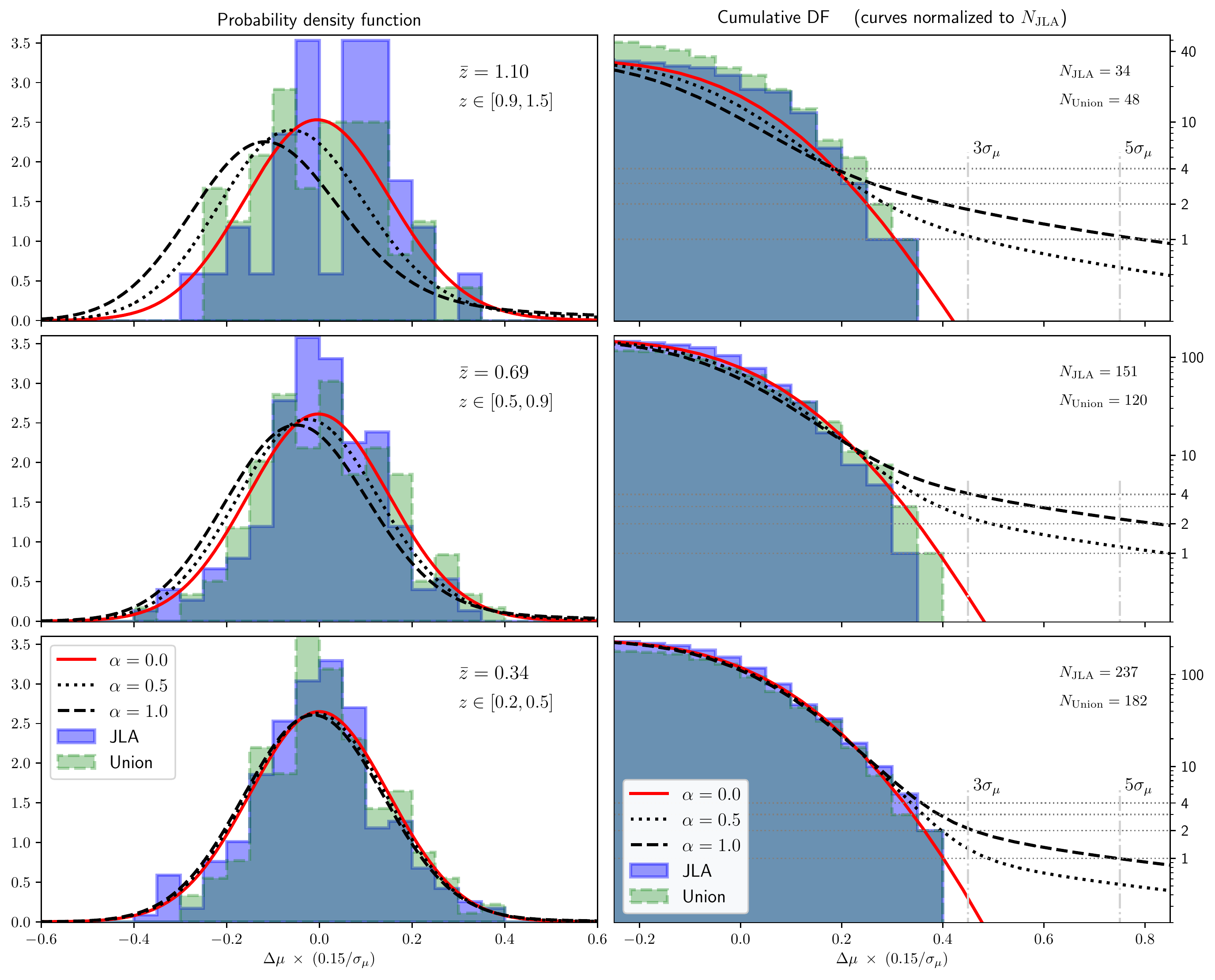}
 \vspace*{-15pt}
 \caption{
 Probability of lensing magnification  $\Delta\mu$ (Eq. \ref{eq:d_lum_magnified}) and its dependence on SNe redshift $z$ and compact object abundance $\alpha ={\Omega_{\rm PBH}}/{\Omega_M}$.
 A sizable compact object population displaces the maximum of the PDF towards the empty-beam distance, compensated with a probability tail for high magnification.
 Both effects grow with redshift, we show only $z\geq 0.2$.
 \\
 \textbf{Left panel:} Probability density function (Eq. \ref{eq:likelihood}), normalized to unity. 
 Histograms show to residuals of JLA (blue, solid) and Union 2.1 (green, dashed) data in the redshift ranges shown. 
 Curves show theoretical predictions for negligible PBH (red, solid), 50\% PBH (black, dotted) and PBH-only (black, dashed) universes at the mean redshift $\bar z$ of the subsample.
 A fiducial Gaussian scatter with typical SNe uncertainties $\sigma_\mu=0.15$ has been assumed to facilitate comparison of theory and data.
 \\
 \textbf{Right panel:} Tail distribution (cumulative) in logarithmic scale to highlight the enhanced  probability of high-magnification in PBH models.
 Histograms are normalized to the number of SNe in each redshift interval and theory predictions are normalized to the number of JLA SNe.
 Horizontal lines correspond to 1-4 events and vertical lines mark where 3$\sigma_\mu$, 5$\sigma_\mu$ outliers are expected, relative to the SNe measurement uncertainty. 
 \label{fig:pdf_data}
 }
\end{figure*}

\section{Magnification by compact objects}

Magnification by gravitational lensing affects the perceived SNe luminosity, which in turn modifies the inferred distance
\begin{equation}\label{eq:d_lum_magnified}
 D_L(z,\Delta \mu) = \frac{\bar D_L(z)}{\sqrt{1+\Delta\mu}}\,.
\end{equation}
Here the magnification $\Delta\mu$ is defined with respect to the average (full beam) luminosity distance  $\bar D_L(z)$.
The probability of a given magnification is described by a probability density function (PDF) that depends in general on redshift, cosmological parameters and the properties and abundance $\alpha\equiv \Omega_{\rm PBH}/\Omega_M$ of compact objects (i.e. PBH) relative to the total matter density. The hypothesis of compact objects encompassing 100\% of dark matter corresponds to $\alpha = \Omega_{CDM}/\Omega_M \approx 0.844$ \cite{Ade:2015xua}.

\begin{figure*}[t!]
\includegraphics[width=0.99\columnwidth]{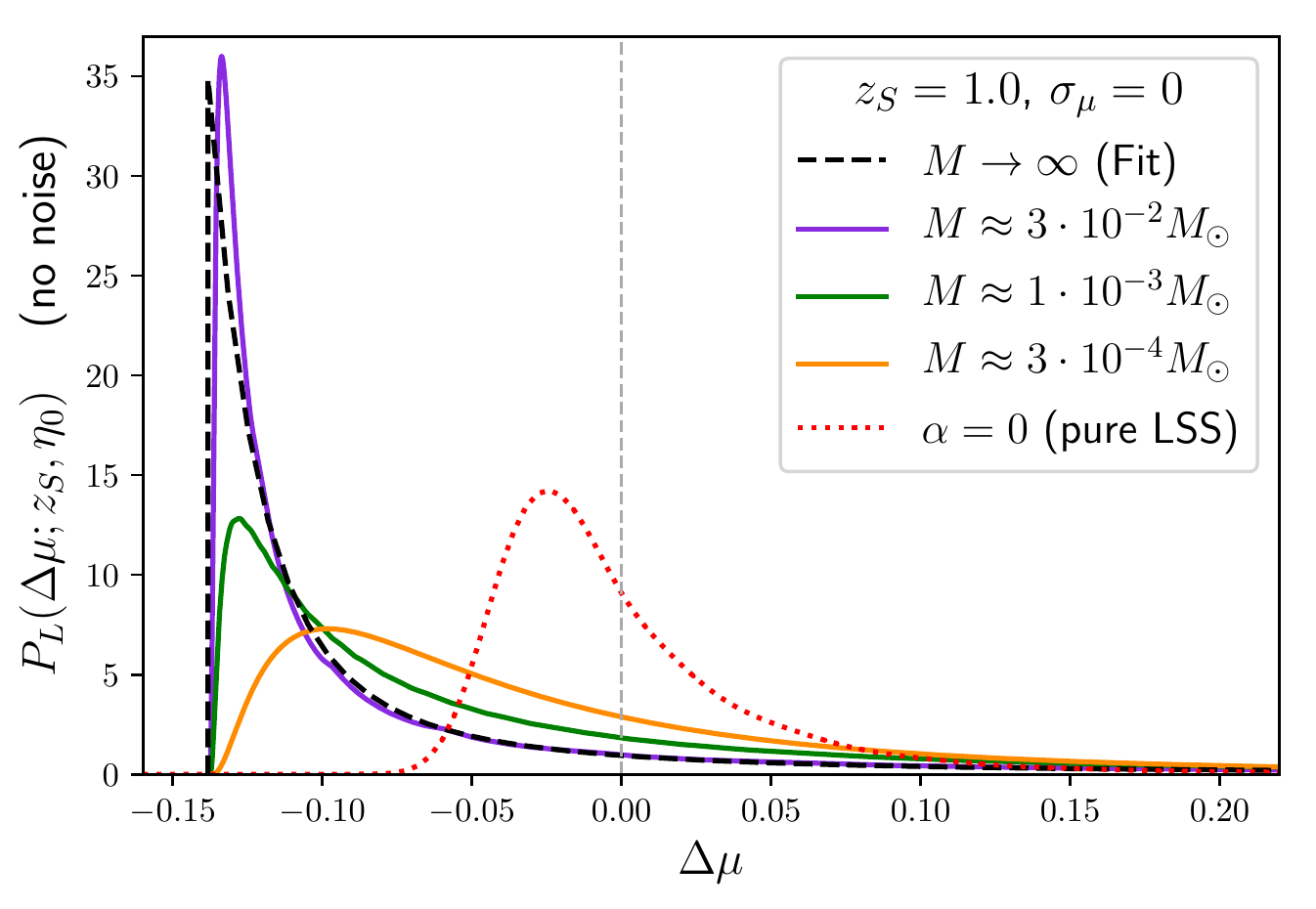}
 \includegraphics[width=0.99\columnwidth]{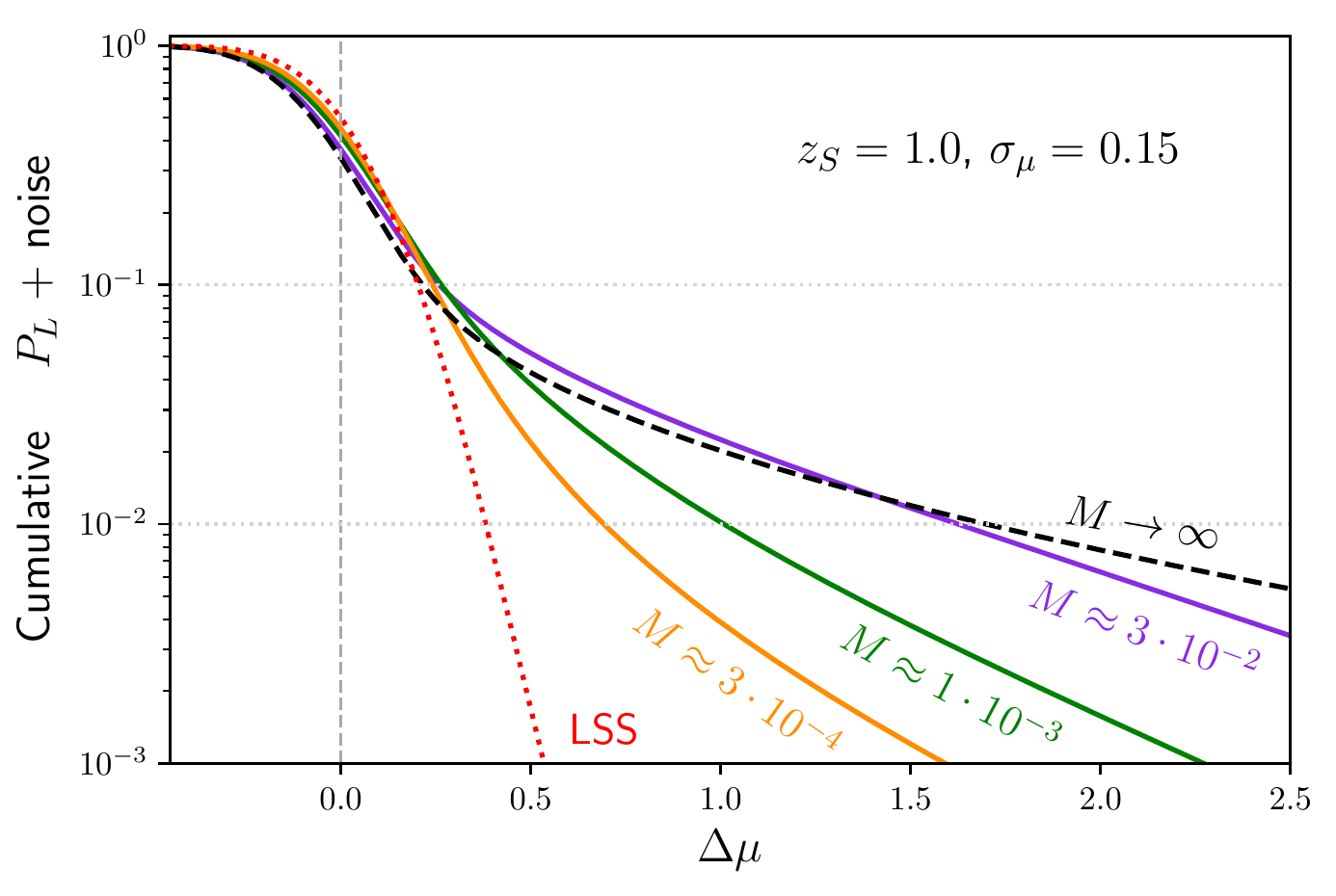}
 \caption{
Magnification PDF dependence on the PBH mass for extended SNe (compare with Fig. \ref{fig:pdf_data}). The curves for each PBH mass assume a SNe radius $R_S \approx 115 AU$ (see Supplemental Material Sec. \ref{sec:finite_sources}). 
\\
\textbf{Right panel:} PDF without noise. For $M\lesssim 10^{-2}M_\odot $ the result converges the analytical fit used in the analysis \cite{Rauch:1991}. Note that introducing realistic noise (as in Fig. \ref{fig:pdf_data}) would render both curves practically indistinguishable around the peak. 
\\
\textbf{Left panel:} cumulative PDF, convolved with noise. The high-magnification tail decays faster than $(\Delta\mu)^{-3}$ but a significant fraction of highly magnified SNe are predicted even for low PBH masses.
 \label{fig:finite_size_pdf}
}
\end{figure*}

%lensing explained
Compact objects induce two characteristic signatures
in the magnification PDF \cite{Rauch:1991,Metcalf:1999qb,Seljak:1999tm,Metcalf:2006ms}:
1) Most objects appear dimmer than the average, as most light bundles do not pass near any lens, shifting the maximum probability towards the empty beam demagnification 
and
2) a few objects undergo significant magnification, as their light bundles pass very close to one or several lenses, appearing as a tail of overluminous outliers in the PDF.
The two effects are shown in Fig. \ref{fig:pdf_data} together with the SNe data used in the analysis (both effects compensate since the average magnification is zero). 
In the limit where lenses and sources can be considered as point-like and the mean magnification is small (so collective lensing effects such as caustic networks are negligible), the PDF is independent of the mass and distribution of compact objects \cite{Rauch:1991,Metcalf:2006ms}. 
We use a magnification PDF that combines the effects of compact objects with the distribution of cosmological LSS \cite{Seljak:1999tm}.
See Supplemental Material Sec. \ref{sec:lensing_pdf}, \ref{sec:lensing_lss} for details on lensing by compact objects and the role of LSS.

%finite sources explained
The point-like source approximation limits SNe lensing bounds to $M_{\rm PBH}\gtrsim 0.01 M_\odot$, for typical Ia SNe sizes \cite{Goldstein:2017}. The criterion is for the SNe size in the lens plane to be small compared to the Einstein radius of the lens. Since the Einstein radius grows with the lens mass, the results converge to the point-source approximation in the limit $M_{\rm PBH}\to \infty$, even for finite SNe. 
We explicitly computed the finite source magnification PDF for different PBH masses \cite{Pei:1993}, finding an excellent agreement with the point-like result for $M_{\rm PBH}\gtrsim 0.03 M_\odot$ across the magnifications relevant for the analysis (see Fig. \ref{fig:finite_size_pdf}). Similar computations for even smaller mass suggest that the PBH signatures might remain competitive even for values as low as $M_{\rm PBH}\sim 3\cdot10^{-4}M_\odot$.
For a numerically efficient treatment, we define an effective lenses fraction that only counts PBHs able to magnify a given SN above a threshold based on the maximum magnification. This prescription is used to derive constraints depending on the PBH mass, as shown in Fig. \ref{fig:mass_dep_constraints}. Our criterion is very conservative given the computation of the full PDF in Fig. \ref{fig:finite_size_pdf}.
See Supplemental Material Sec. \ref{sec:finite_sources} for details on finite-source effects in the PDF and PBH constraints.

\section{Supernovae Analysis}

We will adopt a form of Bayesian hierarchical modeling for our 
statistical analysis and apply it to the Joint Likelihood Analysis (JLA) sample \cite{Betoule:2014frx} and Union 2.1 \cite{Suzuki:2011hu}. The (unobservable) lensing magnification of each SNe is determined by a latent variable $\Delta\mu_i$. Rather than sampling a high-dimensional parameter space including the set $\{\Delta\mu_i\}$, we perform a convolution of the total lensing PDF with the intrinsic error associated to each SNe
\begin{equation}\label{eq:likelihood}
 L_i(\vec\theta,\alpha) = \int d\Delta\mu_i P_{L}(\Delta\mu_i;,z_i,\alpha)P_{SNe}(\Delta m_i,\sigma_i,z_i,\vec \theta)\,.
\end{equation}
Here $P_L$ is the total magnification PDF (described above). $P_{SNe}$ is a general non-gaussian PDF that accounts for the 
intrinsic distribution of SNe luminosities and the observational noise 
(assumed to be a Gaussian with mean zero and variance $\sigma_i$). 
The quantity $\Delta m_i = m_i - \left(5\log_{10}\left( \frac{\bar D_L(z,\Omega_M)}{\text{Mpc}}\right) + 25 - 2.5\log_{10}(1+\Delta \mu)\right)$ is the difference between the (corrected) observed magnitude and the model prediction (equation \ref{eq:d_lum_magnified}), including magnification.
The vector $\vec\theta$ collectively denotes additional parameters describing the cosmology (matter fraction $\Omega_M$) and SNe population (mean magnitude, excess scatter, skewness and kurtosis, respectively $\bar m$, $k_2$, $k_3$, $k_4$). 
For the JLA sample $\vec\theta$ includes, in addition, the SNe standardization parameters (stretch, color and host, respectively $a$, $b$, $\Delta_M$). 
We will assume that the total likelihood is the product of individual likelihoods  $L = \prod_i L_i$ and discuss correlated noise separately. 
We sample the space of parameters spanned by $\alpha,\vec\theta$, assuming a spatially-flat universe with a Gaussian prior on $\Omega_M = 0.309 \pm 0.006$ \cite{Alam:2016hwk} and the remaining cosmological parameters fixed to Planck best fit \cite{Ade:2015xua}.
See Supplemental Material Sec. \ref{sec:sne_likelihood}, \ref{sec:sne_pdf} for details on the likelihood and SNe population modeling.

\begin{figure}[t!]
% \hspace*{-10pt}
 \includegraphics[width=1.0\columnwidth]{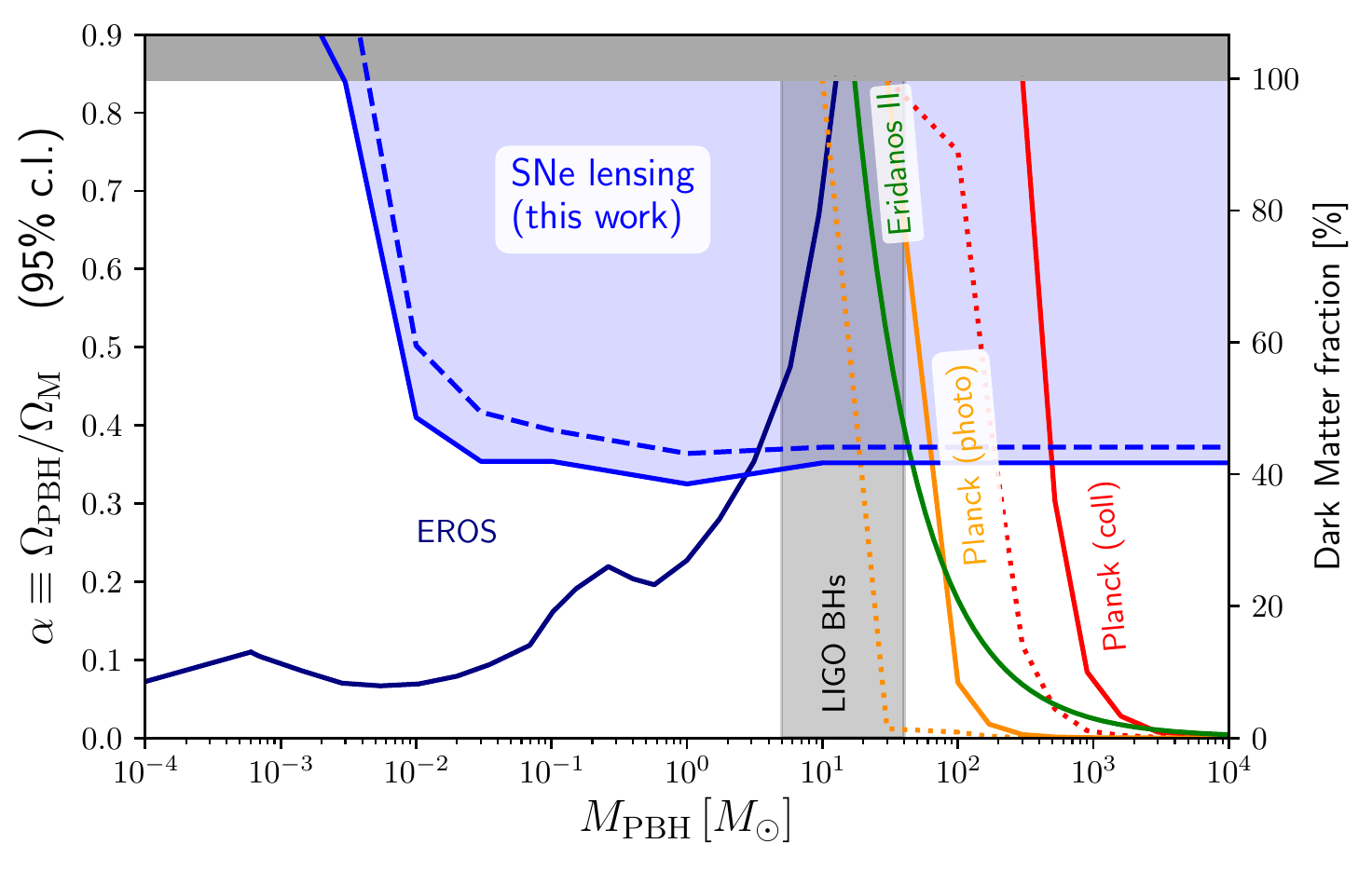}
 \caption{Bounds on the abundance of PBHs as a function of the mass (95 \% confidence level). The analysis of SNe lensing using the JLA (solid) and Union 2.1 compilations (dashed) constrain the PBH fraction in the range $M\gtrsim 0.01 M_\odot$. This range includes the masses of black hole events observed by the Laser Interferometer Gravitational-Wave Observatory (gray), only weakly constrained by previous data including micro-lensing (EROS \cite{Tisserand:2006zx}), the stability of stellar compact systems (Eridanus II \cite{Brandt:2016aco,Li:2016utv}) and CMB \cite{Ali-Haimoud:2016mbv,Bernal:2017vvn}. The CMB excluded regions correspond to Planck-TT (solid), Planck-full (dotted) for the limiting cases of collisional (red) and photo-ionization (orange) (see \cite{Bernal:2017vvn} for details). 
 \label{fig:mass_dep_constraints}
 }
\end{figure}

%main result
Our analysis provides stringent bounds on the compact object abundance, $\alpha< 0.352$ (JLA) and $\alpha<0.372$ (Union) at 95\% confidence-level in the limit $M_{\rm PBH} \gg 0.01 M_\odot$. 
The PBH abundance $\alpha$ is very weakly correlated with the parameters in the SNe population, due mainly to the redshift dependence of the PBH signatures. In our baseline analysis the skewness (for Union) and both skewness and kurtosis are compatible with zero, suggesting that the non-Gaussian SNe distribution used in the likelihood (equation \ref{eq:likelihood}) is sufficiently general. 
Similar analyses fixing the compact-object mass $M_{\rm PBH}$ show how the constraints degrade, as the fraction of effective lenses reduces with decreasing mass (see Fig. \ref{fig:mass_dep_constraints}).
We note that the independence of the PDF to the specific mass distribution of compact objects makes the bounds sensitive to the total fraction in objects with masses $M_{\rm PBH} \gtrsim 0.01 M_\odot$.
The constraints degrade slightly when the Planck+BAO prior on $\Omega_M$ is lifted, leading to $\alpha < 0.440$ (JLA) and $\alpha < 0.437$ (Union) at 95\% c.l. where the difference is due to a degeneracy between the empty-beam shift and the matter fraction. We note that the constraints remain competitive due to the lack of highly-magnified events.
See Supplemental Material Sec. \ref{sec:base_analysis} for the complete discussion of our base analysis.

% Outliers
A potentially important systematic effect is the removal of outliers with large residuals from the base dataset, as overluminous SNe could be either intrinsically brighter (e.g. peculiar classes, contamination) \cite{Taubenberger:2017hoo} or highly-magnified events (e.g. due to PBHs). To address this issue we used the outliers rejected from the Union sample, noting however that most of those outliers have features that suggest that they are peculiar SNe rather than due to gravitational magnification (5/8 underluminous and 3/4 overluminous).  
Including all the outliers from the Union sample degrades the constraints slightly to $\alpha < 0.413$ (95\% c.l.). This is due to the larger abundance and significant deviations of underluminous outliers, which is better fit by a non-zero kurtosis parameter.
Considering only overluminous outliers (as preferred by compact-object models, cf. Fig. \ref{fig:pdf_data}) still results in bounds $\alpha < 0.573$ (95\% c.l.). This is due to data not agreeing with the maximum magnification probability being around the empty-beam demagnification.
See Supplemental Material Sec. \ref{sec:outliers} for the discussion of SNe outliers.

%other systematics
Additional analyses allowed us to establish the robustness of our results against systematic effects.
%correlations and selection
We studied the impact of correlated noise using model based on the compressed JLA likelihood with an additional free parameter. Our prescription shows that correlated noise does not have a strong effect, as it leads to only a 5\% modification of the base JLA results ($\alpha < 0.363$ at 95\% c.l.). 
Selection bias is less problematic than in standard cosmological analysis with broad priors in $\Omega_M$, as the differences due to cosmology are larger than those caused by lensing. 
%SNe evolution
SNe population evolution across redshift may weaken the bounds similarly to lifting cosmological priors but can not explain the lack of highly-magnified SNe.
See Supplemental Material Sec. \ref{sec:sne_covariance}, \ref{sec:selection_bias} for the discussion of correlated noise, selection effects and SNe evolution.

\section{Conclusions}

%outlook
Our results on the compact object abundance reject the hypothesis of dark matter being entirely composed of stellar-mass primordial black holes at the level of 4.79$\sigma$ (JLA) and 4.54$\sigma$ (Union). 
The significance of the exclusion remains at the level of 2.90$\sigma$ (Union) when interpreting overluminous outliers as magnified SNe (despite indications that 3/4 of such events are peculiar SNe). 
Primordial black holes need to be light ($M_{\rm PBH}\lesssim 0.01 M_\odot$ and hence subject to stellar microlensing bounds) or a sub-dominant contribution to the dark matter.  Note that an extended mass function only lowers the constraints if the majority of the total mass is in the form of light PBH.

SNe constraints fully cover the mass range of LIGO events and supplement other tests of macroscopic dark matter (see Fig. \ref{fig:mass_dep_constraints}).
Our analysis is complementary to stellar microlensing \cite{Tisserand:2006zx}, which relies on the real-time evolution of the magnification and thus less sensitive in the limit of high PBH mass. In contrast, SNe lensing relies on the known luminosity rather than on the relative motion of lens and source, and is thus effective in the opposite limit of heavy lenses, where large Einstein radii make it more likely to produce highly magnified objects.  
Our results on the PBH fraction agree with recent constraints based on microlensing of quasars \cite{Mediavilla:2017bok} and stars \cite{2011MNRAS.413..493W}, caustic crossings \cite{Diego:2017drh,Venumadhav:2017pps,Oguri:2017ock}, as well as revised estimates of LIGO event rates \cite{Sasaki:2016jop,Ali-Haimoud:2017rtz,Chen:2018czv,Magee:2018opb,Abbott:2018oah}, radio and X-ray emission \cite{Gaggero:2016dpq}, 21-cm absorption \cite{Hektor:2018qqw}, pulsar timing arrays \cite{Schutz:2016khr} and the less conservative bounds from the cosmic microwave background \cite{Ali-Haimoud:2016mbv,Bernal:2017vvn,Poulin:2017bwe}. 
Our constraints translate directly to other compact objects with $M\gtrsim 0.01M_\odot$, e.g. \cite{Ricotti:2009bs,Aslanyan:2015hmi}.

Our analysis improves substantially on previous SNe lensing studies \cite{Metcalf:2006ms}, reflecting the evolution of the quality and quantity of data. Larger  SNe catalogues (e.g. \cite{Scolnic:2017caz,Jones:2017udy,Kessler:2016uwi}) will significantly increase the constraining power of this technique \cite{Dhawan:2017kft}.
Gravitational lensing methods together with a variety of other techniques involving gravitational waves \cite{Clesse:2016ajp,Kovetz:2016kpi,Cholis:2016kqi,Raidal:2017mfl,Ali-Haimoud:2017rtz,Magee:2018opb}, lensing of fast radio bursts \cite{Munoz:2016tmg}, astrometry \cite{VanTilburg:2018ykj}, pulsar timing \cite{Schutz:2016khr} and CMB \cite{Ali-Haimoud:2016mbv,Horowitz:2016lib, Poulin:2017bwe, Bernal:2017vvn,Nakama:2017xvq} (among others) hold considerable promise to constrain the abundance and properties of primordial black holes at increasing significance.

\acknowledgments{
We are very grateful to D. Rubin for conversations and providing us with the Union outlier sample, L. Galbany for an appraisal of these ouliers and the authors of Ref. \cite{Bernal:2017vvn} for the CMB data in Fig. \ref{fig:mass_dep_constraints}, as well as G. Aldering, T. Collett, D. Goldstein, J. Guy, B. Hayden, D. Holz, T. de Jaeger, D. Jones, D. Kasen, A. Kim, E. Mediavilla, B. Metcalf, D. Scolnic and  S. Perlmutter for useful discussions. 
MZ is supported by the Marie Sklodowska-Curie Global Fellowship Project NLO-CO. This research used resources of the National Energy Research Scientific Computing Center, a DOE Office of Science User Facility supported by the Office of Science of the U.S. Department of Energy under Contract No. DE-AC02-05CH11231
}

\setcounter{figure}{0}
\setcounter{table}{0}
\renewcommand{\thefigure}{S\arabic{figure}}
\renewcommand{\thetable}{S\arabic{table}}
\appendix

\begin{center}
\phantom{a}
\vspace{15pt}
{ \it \large Supplementary Material}\\ 
\end{center}

% \tableofcontents

\section{SNe Lensing by Compact Objects} \label{sec:lensing}

%Global element
Gravitational lensing of small sources like SNe is sensitive to the abundance of compact objects. This section presents the statistical predictions of lensing magnification, including the effects of a variable PBH fraction and the large scale structure (LSS) of the universe. We will then consider how the constraints are affected by assumptions on the PBH mass distribution due to the finite SNe size.

\subsection{Magnification by compact objects}\label{sec:lensing_pdf}

\begin{figure*}%[t!]
 \includegraphics[width=0.49\textwidth]{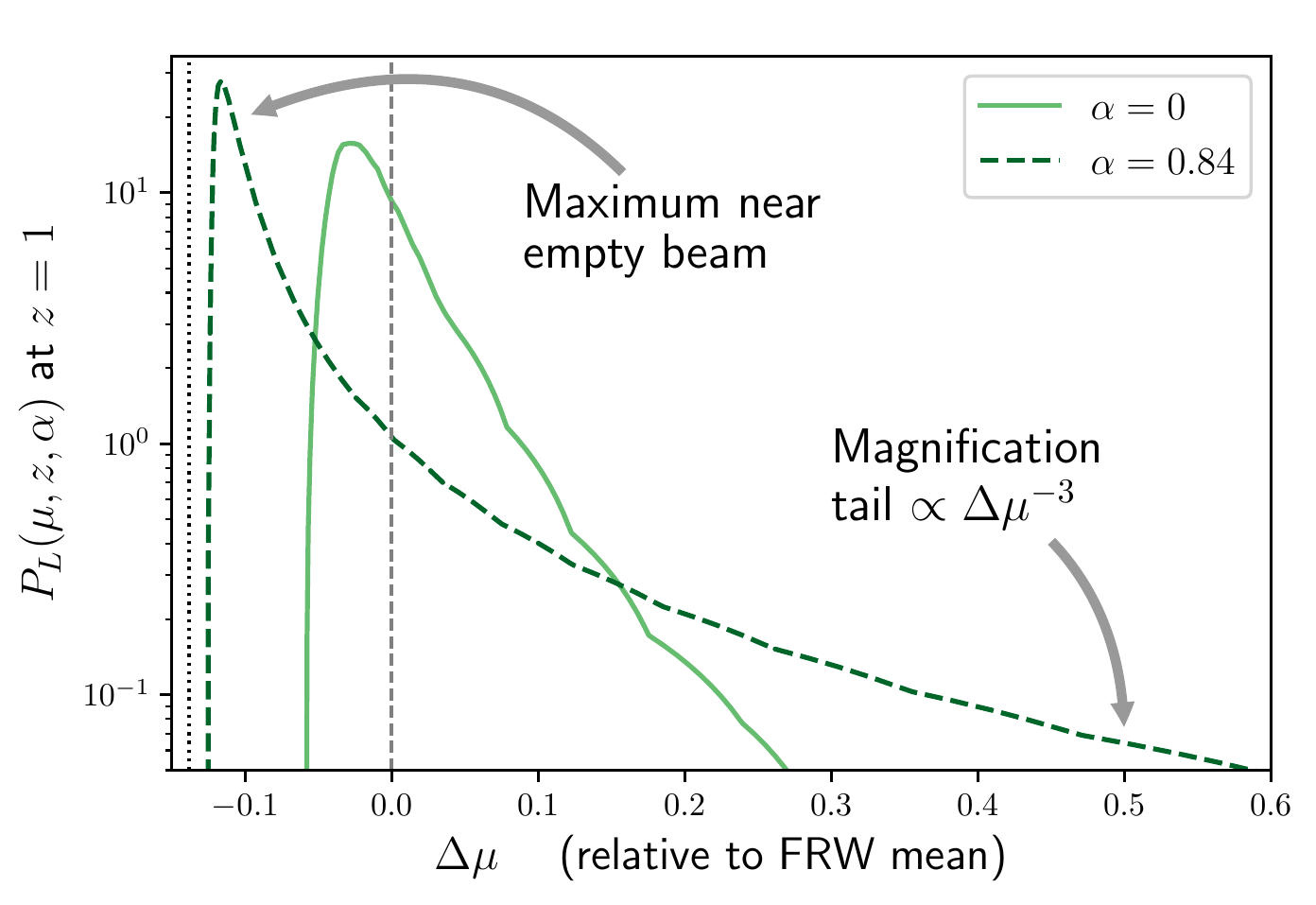}
  \includegraphics[width=0.49\textwidth]{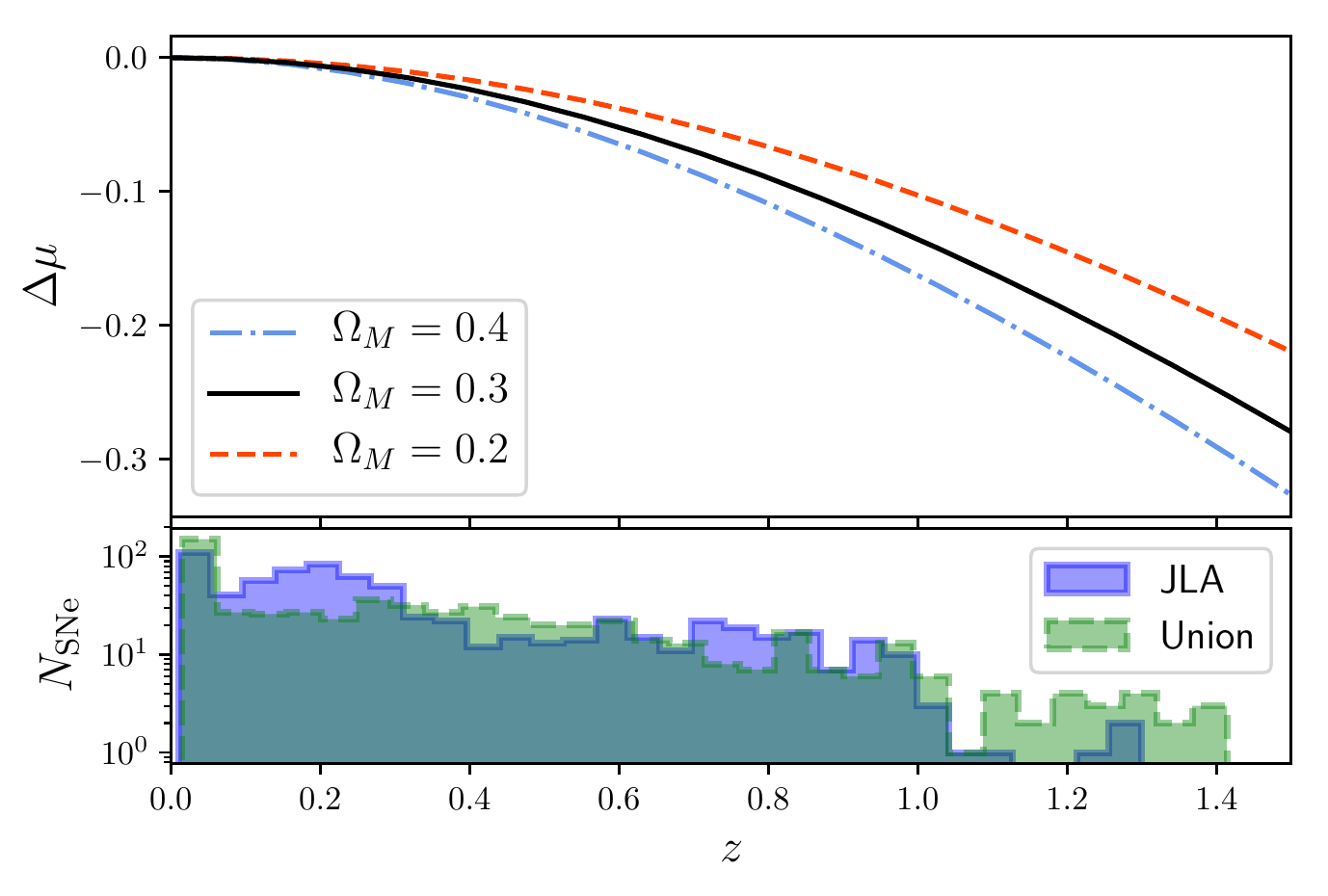}
 \caption{
Lensing in a universe with compact objects. \textbf{Right panel:}
Effects of the PBH fraction on the magnification probability density function (equation \ref{eq:p_theory}), including compact objects and cosmological large scale structure. Compact objects produce 1) a displacement of the maximum of the PDF towards a demagnified universe and 2) a larger probability of large magnifications. The cases shown correspond to no PBH (solid) and all of the dark matter (but not baryons) in PBH (dashed) at $z=1$. Also shown is the 
empty beam (vertical dotted line). We see that the probability of reaching empty beam values is negligible for LSS.
\textbf{Left panel:} redshift dependence of the shift in the peak of the magnification PDF (equation \ref{eq:friedman_mu}) for different cosmologies, along with the SNe distribution. The effect is stronger at high redshifts where there are fewer supernovae (lower panel). 
 \label{fig:pbh_signatures}
 }
\end{figure*}

%distance and magnification, empty beam
In PBHs-DM universe, the line of sight to most sources will not pass near any compact object. Those sources will be demagnified and appear fainter, affecting its perceived angular-diameter distance
\begin{equation}\label{eq:magnif_rel_empty}
D(\mu,z) = \frac{\bar{D}(z)}{\sqrt{1+\Delta\mu}} =\frac{D_{E}(z)}{\sqrt{1+\mu}}\,.
\end{equation}
In the first equality we have defined the magnification $\Delta\mu$ relative to \emph{filled-beam} distance, i.e. the angular diameter distance of the homogeneous cosmology $\bar D(z) = \frac{1}{1+z}\int \frac{dz^\prime}{H(z^\prime )}$. 
The second equality defines $\mu$ relative to the \emph{empty-beam} distance \cite{Dyer:1973zz,Das:2005yb}
\begin{equation}\label{eq:empty_beam}
 D_{E}(z) = \int_0^z dz^\prime \frac{1}{(1+z^\prime)^2 H(z^\prime)}\,. %\frac{1}{g(z^\prime)} 
\end{equation}
This is shown in Fig. \ref{fig:pbh_signatures}.

%PDF for compact objects and high magnification
Some sources will appear highly magnified by a compact object near the line of sight. 
The lensing probability distribution function (PDF) of a universe filled by a uniform comoving density of PBHs only depends on the mean magnification $\bar\mu$. Numerical simulations \cite{Rauch:1991} have found that that the PDF is well approximated by 
\begin{equation}\label{eq:pdf_compact}
 P_C(\mu,\bar\mu) = 
 A\left[\frac{1-e^{-\mu/\delta}}{(\mu+1)^2-1}\right]^{3/2} \,\quad \text{ if } \mu>0 \,,
\end{equation}
and $0$ otherwise. The parameters $A$, $\delta$ depend on $\bar\mu$ and are chosen to normalize the distribution and enforce the mean $\bar\mu \equiv \int d\mu \, \mu P_C(\mu,\delta)$ (note that $\bar\mu$ depends on redshift in general). 
In the high-magnification limit the PDF decays as $P_C(\mu)\propto \mu^{-3}$, as has been shown in the limit of a single lens \cite{Metcalf:2006ms} and by detailed numerical studies with a distribution of point lenses \cite{Rauch:1991,Holz:1997ic}.
We note that our analysis is in the regime of low optical depth (low average convergence and shear): we will use data with $z \sim 1$ where $\bar{\mu} < 0.14$ and the PDF for point lenses only depends on the total magnification \cite{Rauch:1991,Metcalf:2006ms}. In this regime 
caustics are isolated and from individual lenses only and magnification is well below the threshold where collective effects (caustic networks) become important. 
In this limit the PDF (equation \ref{eq:pdf_compact}) is independent of the PBH mass as long as lenses and sources can be considered point-like, with the finiteness of sources requiring that $M_{\rm PBH}\gtrsim 10^{-2}M_\odot$ (see Sec. \ref{sec:finite_sources}).

%transition to next section
In the specific case of a PBH-only universe the mean magnification in the PDF (equation \ref{eq:pdf_compact}) has to ensure that the mean distance corresponds to the homogeneous cosmology ($\Delta\mu = 0$ in equation \ref{eq:magnif_rel_empty}). This corresponds to $\bar\mu =  \mu_F $  where the full-beam magnification is
\begin{equation}\label{eq:friedman_mu}
 \bar{\mu} \equiv \left({D_{E}(z)}/{\bar D(z)}\right)^2-1\,.
\end{equation}
Flux conservation demands that the mean magnification remains unchanged 
regardless of the nature of perturbations. 
Allowing $\bar\mu$ to vary along individual directions we can also include the effects of LSS clustering in the next section, but even then the mean
magnification averaged over all lines of sight has to be given by this 
expression. 

\subsection{PBH fraction and Large Scale Structure}  \label{sec:lensing_lss}

%PBH fraction
We need to generalize the simplified model of the previous section to a realistic universe in which a fraction of compact objects 
\begin{equation}
 \alpha \equiv \frac{\rho_{\rm PBH}}{\rho_M} = f_{\rm PBH} \frac{\rho_M}{\rho_{DM}} \,,
\end{equation}
traces the underlying LSS distribution. Note we are distinguishing the fraction over the total matter density including baryons $\alpha$ from the fraction of DM $f_{\rm PBH}$. We will work with the former, as it is more convenient to incorporate the effects of LSS.

%convolution with LSS
Lensing by compact objects (equation \ref{eq:pdf_compact}) and LSS  will each contribute with a weight depending on the PBH fraction $\alpha$.
For a given line of sight, the mean magnification is that associated with the LSS distribution, regardless of $\alpha$. In the absence of compact objects 
this LSS PDF is shown in Fig. \ref{fig:pbh_signatures}: one can see 
that PDF samples $\mu^\prime$ values peaked around 0 (mean beam), with a tail of 
rare events towards high magnifications caused by matter in centers of 
halos (galaxies and clusters). In the presence of compact objects 
a fraction $\alpha$ of LSS magnification $\mu^\prime$ along each line of sight is spread out further with its own PBH PDF: this is constructed such that it conserves 
the mean magnification $\alpha \mu^\prime$ of that line of sight (determined by LSS PDF), but its distribution is more peaked at 
empty beam, with a tail towards high magnifications. 
The total magnification $\mu$ is then obtained by adding 
a contribution $(1-\alpha)\mu^\prime$ to the  
contribution from compact objects, where the compact objects PDF has a mean magnification $\alpha\mu^\prime$ \cite{Seljak:1999tm}. 
This approach yields a combined lensing PDF
\begin{equation}\label{eq:p_theory}
  P_L(\mu;z,\alpha) = \int_0^{\frac{\mu}{1-\alpha}} d\mu^\prime P_{LSS}(\mu^\prime,z)P_C[\mu-\mu^\prime(1-\alpha),\alpha\mu^\prime]
\end{equation}
where $P_{LSS}(\mu,z)$ is the PDF associated to LSS.
We see that the result is a 
convolution of the two PDFs. 

We take $P_{LSS}$ from turboGL \cite{Kainulainen:2009dw,Kainulainen:2010at} for a Planck cosmology \cite{Ade:2015xua}. This code computes the PDF of LSS using 
the halo approach, considering halos with mass $M> 10 M_\odot /h$ and a Navarro-Frenk-White profile.
The halo modeling should be more accurate than simulations \cite{Seljak:1999tm} due to the high 
dynamic range of the halo mass profile resolution needed. We note however that around the peak there is an excellent agreement between the different LSS PDFs in the literature.
Still, there are effects that are model dependent in the centers of the halos, such as the stellar and 
baryonic contribution, which affect the rare event tail of LSS PDF. 

\begin{figure*}%[t!]
 \includegraphics[width=0.48\textwidth]{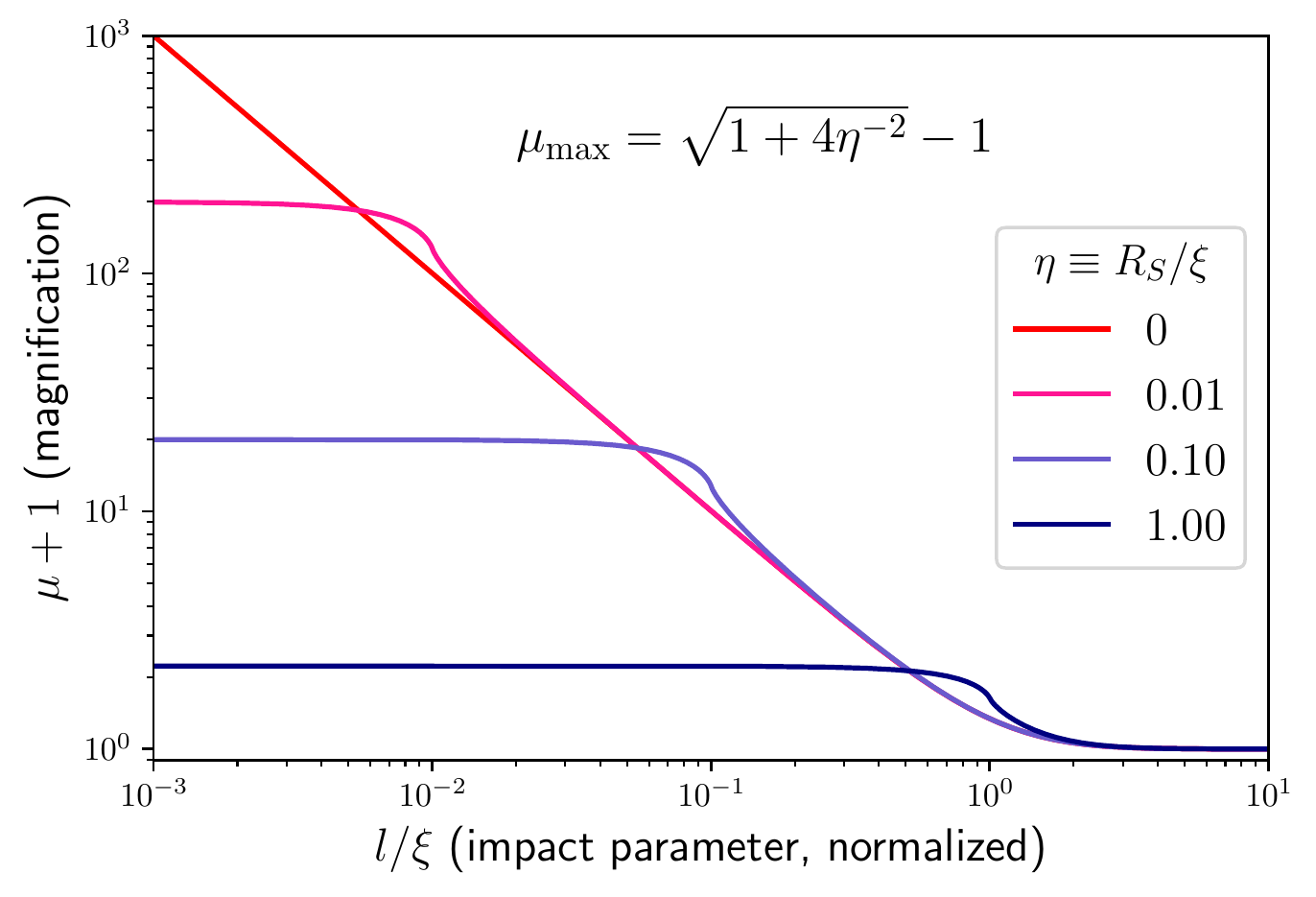}
 \includegraphics[width=0.49\textwidth]{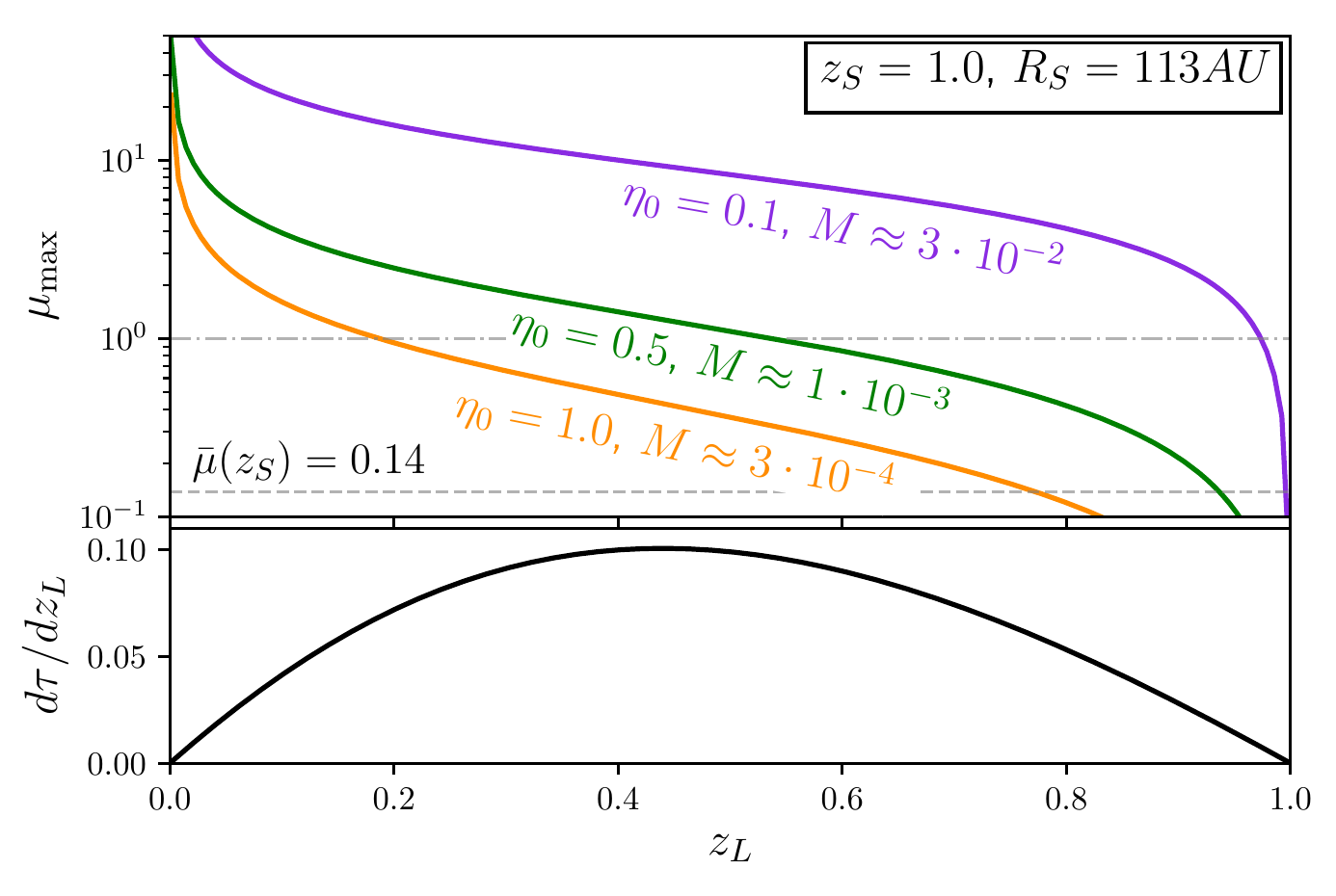}
 \caption{
Breakdown of the point-source approximation for type Ia supernovae.
\textbf{Right panel:} Magnification as a function of the impact parameter \cite{Schneider87}, normalized to the Einstein radius. The magnification saturates for $l/\xi\lesssim \eta$ for low impact parameter values $l\lesssim \eta$.
\textbf{Left panel:} Maximum magnification of a source at $z_S = 1$ by a lens at $z_L$ (upper panel) for different values of the size parameter $\eta_0$ (corresponding lens mass in units of $M_\odot$). Horizontal lines show $\mu=1$ and $\bar\mu$ (equation \ref{eq:friedman_mu}), showing that SNe can be highly magnified even for very light PBHs, particularly at low redshift. 
The differential optical depth (lower panel) weights the effective contribution of lenses at different redshift $z_L$.
 \label{fig:finite_size_magnification}
}
\end{figure*}

So far there have been only a handful of SN detected that have been strongly magnified. All of 
these events are consistent with the lensing effect generated by lenses that have been identified as galaxies or clusters \cite{Quimby:2014irr,2017Goobar,2017Rubin} and the magnification is fully consistent with the lens properties.
The situation would be very different if these SNe were magnified by PBHs: a signature of the PBH model is that the lenses can not be identified.
We also note that some observed magnified SNe are a result of a targeted search towards special objects such as clusters, and their probability density cannot easily be translated onto our plot. Here we do not consider those events in which the SN has passed through a center of a halo, since the modeling uncertainties are too large to use these events to distinguish between the LSS lensing and compact object lensing. 
Note that the effect of the cosmological parameters on the lensing PDF is very weak \cite{Castro:2014oja,Macaulay:2016uwy} and variations can be neglected in the range allowed by Planck.  The combined PDF preserves both the maximum at the empty-beam distance and the high magnification tail (see Fig. \ref{fig:pdf_data} in the letter).

\subsection{Finite sources and PBH mass distribution} \label{sec:finite_sources}

%considering small masses
The magnification PDF relies on the assumption that sources and lenses are point-like. Beyond that regime, one needs to take into account the finite size of the source.
To quantify whether a given SNe is an effective point-like source for a point-like lens of mass $M$ we take the ratio of the SN size and the Einstein radius $\xi$ 
\begin{equation}
 \eta(z_L,z_S) = \frac{R}{\xi} = \eta_0\left[\frac{D_L}{H_0 D_S D_{LS}}\right]^{1/2} \,,
\end{equation}
where $\xi$ is the Einstein radius of the system and $R$ is the radius of the source in the lens plane, related to $R_S$, the physical size of the source, by $R/D_L=R_S/D_S$. The dimensionless  effective size parameter reads \cite{Pei:1993}
\begin{equation} \label{eq:point_like_ratio} 
\eta_0= 0.0235 \sqrt{h}
\left( \frac{R_{S}}{115.5 {AU}} \right)
\sqrt{\frac{M_{\odot}}{M}}
 \,,
\end{equation}
Models of Ia SNe find that 
intensity weighted expansion velocity is of order 10,000km/s \cite{Goldstein:2017}, with the
peak magnitude reached after 20 days, leading to a typical SN Ia size of 
$1.73 \cdot 10^{10}$km ($115.5 AU$), a value we will adopt here.

The magnification by a single lens depends on the transverse separation between the lens and the source $l$. The point-like approximation is valid  if $l/\xi$ is larger 
than the source size $\eta$ (see Fig. \ref{fig:finite_size_magnification}, left panel). The finite size of the source introduces a maximum magnification $\mu\leq \mu_{\rm max} = \sqrt{1+4/\eta^2}-1$, with equality in the limit of perfect alignment ($l=0$). The maximum magnification depends on redshift, with a fraction of low redshift lenses typically able to produce $\mathcal{O}(1)$ magnifications and larger (see Fig. \ref{fig:finite_size_magnification}, right panel). 
The effect of each lens is then weighted by the \textit{differential optical depth} \cite{1983ApJ...267..488V}
\begin{equation}\label{eq:diff_optical_depth}
 \frac{d\tau}{d z_L} = \frac{3}{2}\alpha\Omega_M H_0^2 \frac{(1+z_L)^2}{H(z_L)}\frac{D_L D_{LS}}{D_S} \,.
\end{equation}
Note that the mass of the lenses factors out of the optical depth, which only depends on the total mass of lenses $\alpha\Omega_M H_0^2$.

%full PDF
For a more detailed analysis, we evaluate PBH PDFs 
as a function of source size using the expressions given in Ref. \cite{Pei:1993}, and shown in Fig. \ref{fig:finite_size_pdf} of the letter for $\eta_0 = 0.1,0.5,1.0$, respectively $M/M_\odot = 0.03,0.001,0.0003 $ (more details will be presented separately).
The point size approximation breaks down first for large magnifications where the transverse 
separation is smallest, leading to a suppression of PDF at high magnifications, but not changing low magnification PDF which remains 
peaked close to the empty beam. As we reduce the PBH mass the Einstein radius becomes 
smaller and $\eta_0$ larger, leading to a suppression of PDF at lower
magnifications (see Fig. \ref{fig:finite_size_pdf} in main article). Eventually, as the Einstein radius 
becomes comparable to SNe size, even the PDF at small magnifications 
gets modified, leading to a shift of PDF peak from empty beam to 
filled beam, and an overall reduction of variance. This transition occurs 
when $\eta_0 \sim 0.5-1$ \cite{Pei:1993}, which corresponds to $M \sim 10^{-3}M_{\odot}$. 
The effects of finite size can be observed also in the high-magnification tail of the distribution, which drops faster for lower PBH masses.
In the limit of large $\eta_0$ the variance vanishes and there is no lensing effect from PBH, only from LSS.

% mass dependence choice
The full PDF for finite SNe is computationally expensive, and thus we turn to a simple model to account for the mass dependence: a BH contributes to $\alpha$ only if $\mu_{\rm max}(z_L,z_S)> \mu_{\rm cut}$, where we set $\mu_{\rm cut} = 1$. 
This prescription is justified by the saturation effect observed in the left panel of Fig. \ref{fig:finite_size_magnification}: for a given lens and source, the probability of a magnification $\mu\geq \mu_{\rm cut}$ is given by the cross section $2\pi l^2\big|_{\mu=\mu_{\rm cut}}$ and is approximately independent of the source size as long as $\mu_{\rm max} \geq \mu_{\rm cut}$.
Setting $\mu_{\rm cut} = 1$ is fairly conservative choice given that no data points fall in the region $\mu\gtrsim 1$ (see Fig. \ref{fig:pdf_data} of the letter).
One can then derive the \textit{effective PBH fraction}
\begin{equation}\label{eq:alpha_eff}
\tilde\alpha(z_S,M) = \alpha \cdot f_{L}(z_S,M)\,,
\end{equation}
(now a function of the source redshift and the PBH mass) in terms of the \textit{effective lenses fraction} %$f_{L}(z_S,M)$
\begin{equation}\label{eq:bh_eff_fraction}
f_L = \frac{1}{\tau(z_S)}\int_0^{z_S} \frac{d\tau}{dz_L}\Theta\left(\mu_{\rm max}(z_L,z_S,M) - \mu_{\rm cut}\right)dz_L \,,
\end{equation}
were each lens is weighted by the differential optical depth (equation \ref{eq:diff_optical_depth}, see Fig. \ref{fig:finite_size_magnification}) and the expression is normalized to the total optical depth  $\tau(z_S) = \int_0^{z_S}\frac{d\tau}{dz_L}dz_L$.
Here $\Theta(x)$ is a step function, ensuring that we only count lenses able to produce magnification larger than $\mu_{\rm cut}$ (a smooth function can be used, but the results depend very weakly on this choice).
We will take $\mu_{\rm cut} = 1$, a reasonable observation given that all SNe data satisfy $\Delta\mu \lesssim 0.5$ (see Fig. \ref{fig:pdf_data} of the letter). This choice degrades the constraints for $M\lesssim 0.01 M_\odot$, which is fairly conservative given that the full magnification PDF for finite sources for $M\sim 0.03 M_\odot$ reproduces the $M\to \infty$ results over the range explored by the data, extending even up to $\Delta\mu\sim 2$ (see Fig. \ref{fig:finite_size_pdf} in letter).

%mass distributions & clustering
Realistic PBH models will be characterized by a mass distribution that accounts for their initial generation and subsequent evolution (see Ref. \cite{Byrnes:2018clq} for a physically motivated PBH mass function).
An extended mass function  $P(M)$ can be easily incorporated in the framework of the effective lenses fraction, generalizing the effective PBH fraction to
\begin{equation}\label{eq:bh_eff_distribution}
 \bar\alpha(z_S) =\int P(M) \tilde \alpha(M,z_S) dM \,,
\end{equation}
with $P(M)$ being the normalized mass probability distribution. 
As expected, in the limit of point sources the lensing PDF is insensitive to the masses and small-scale clustering (as long as collective lensing effects can be neglected) properties of the PBH population \cite{Holz:1997ic}.
For these reasons we will assume a monochromatic mass function in what  follows, but generalizing to an extended mass function is 
straightforward.
Since SNe lensing constraints are independent of the mass spectrum and spatial distribution (as long as collective lensing effects can be ignored), they offer a bound on the total PBH fraction above $10^{-2} M_\odot$.

\section{Lensing Likelihood for Type Ia SNe} \label{sec:likelihood}

The PBHs fraction $\alpha$ affects the observed luminosity of type Ia supernovae via gravitational lensing. In this section we present the likelihood used to constrain the model, as it will be applied to the Joint Lightcurve Analisis (JLA) \cite{Betoule:2014frx} and Union 2.1 \cite{Suzuki:2011hu} datasets. The discussion includes the effects of lensing, a general SNe intrinsic luminosity distribution and standardization, as well as potential sources of systematic errors such as outliers, correlated noise and selection bias.

\subsection{Magnification and global likelihood} \label{sec:sne_likelihood}
%Prediction/theory
The effect of compact objects is to change the apparent distance of SNe by $ D_L(z,\Delta \mu) = {\bar D_L(z)}/{\sqrt{1+\Delta\mu}}$ (equation \ref{eq:magnif_rel_empty})
where $\bar D_L(z)=(1+z)^2\bar D(z)$ is the average/full beam luminosity distance (cf. equation \ref{eq:magnif_rel_empty}) and the magnification $\Delta\mu$ is now defined with respect to the average distance $\bar D$.
The distance modulus of observed SNe then reads
\begin{equation}\label{eq:m_th}
 m_{th} = 5\log_{10}\left( \frac{\bar D_L(z)}{\text{Mpc}}\right) + 25 - 2.5\log_{10}(1+\Delta \mu) \,.
\end{equation}
The probability of a given magnification $\Delta\mu$ is given by $P_L(\mu,z,\alpha)$  (equation \ref{eq:p_theory}) evaluated at $ \mu = \Delta\mu + \mu_F$ (equation \ref{eq:friedman_mu}), and using the effective PBH fraction $\alpha\to\tilde\alpha(z_S,M)$ (equation \ref{eq:alpha_eff}) in the case of finite PBH mass.

We will adopt a form of Bayesian hierarchical modeling approach for 
statistical analysis. 
We assume the priors on all parameters are flat, so 
that the posterior is proportional to the likelihood only. 
We introduce for each SNe a 
latent variable which is the true distance modulus, 
which we do not observe, and instead we observe its noisy 
version constructed from absolute magnitude, color and stretch. 
The regression parameters that correct for stretch and color are assumed to 
be the same across all SN. This can be generalized in the 
hierarchical models to allow each SN to have its own value for magnitude, 
color, and stretch, each controlled by a prior with its
own hyper-parameters that can be determined from the analysis itself \cite{2016Nielsen}. However, it has 
been argued this approach is prone to selection bias effects \cite{2016Rubin} and we do not pursue it here. 

There are two general approaches one can follow to solve this: first, one can simultaneously derive the posteriors of all the latent variables and the parameters of the model, which requires an analysis of posteriors in a high dimensional space, for example using Hamiltonian Monte Carlo sampling \cite{Feeney:2017}. The second approach is to analytically marginalize over the latent variables. This requires computing the convolution integrals of the true lensing
probability distribution with the noise probability 
distribution. These integrals cannot be solved in closed-form, and so
need to be done numerically, separately for each 
model, so it needs to be varied over $\alpha$, $z$, and 
intrinsic parameters of the SNe PDF, as well as for each noise level. In this paper we adopt this 
analytic marginalization approach, which requires us to numerically 
compute many convolution integrals and interpolate 
between them. Note that this is a second convolution on top of the one 
between LSS and PBH discussed above. 
However, as a consequence of these analytic 
marginalizations over latent variables we can work with a 
handful of variables only. 

The likelihood for each SNe measurement is then a convolution of the total lensing PDF with the PDF associated with 
a given SNe, which in itself is a convolution of intrinsic SNe PDF 
and observational noise PDF, 
\begin{equation}\label{eq:like_tot}
 L_i(\vec\theta,\alpha) = \int d\mu P_{L}(\mu;,z_i,\alpha)P_{SNe}(m_i,\sigma_i,z_i,\mu,\vec \theta)\,,
\end{equation}
where $m_i$ is the observed distance modulus (equation \ref{eq:m_th}), $\sigma_i$ the corresponding error and $ \mu = \Delta\mu + \mu_F$ (equation \ref{eq:m_th}). The vector $\vec\theta$ collectively denotes parameters describing the cosmology, as well as the standardization and statistics of the SNe population (Sec. \ref{sec:sne_pdf}).

We will assume that the SNe are independent and hence the total likelihood $L = \prod_i L_i$ is the product of the individual likelihood for each SNe. This is a very reasonable assumption for lensing, as correlations induced by compact objects occur on very small angular scales $\sim \theta_E$, and the SNe are observed in random points in the sky. Observational covariances in the SNe datasets due to common 
modeling and systematics are important. The likelihood (equation \ref{eq:like_tot}) is non-gaussian and thus the covariance matrix for the samples can not be straightforwardly included. We will discuss how to 
model the covariance in Sec. \ref{sec:sne_covariance}.

\subsection{SNe standardization, population and errors}\label{sec:sne_pdf}

We want to allow for a sufficiently general, non-gaussian $P_{SNe}$ likelihood in (equation \ref{eq:like_tot}) that can accommodate the distribution of intrinsic luminosities of type Ia SNe.
This distribution will be a function of the normalized deviation between the prediction (equation \ref{eq:m_th}) and the observation:
\begin{equation}\label{eq:norm_deviation}
 x_i = \frac{1}{\sigma_i}\left(m_{ob,i} - m_{th}(z_i,\mu) - \bar m \right)\,.
\end{equation}
Here $\bar m$ controls the mean of the SNe intrinsic magnitude (the observed magnitude $m_{ob,i}$ and corrected dispersion $\sigma_i$  are discussed below).

Since one signature of the signal we are searching for 
is the non-gaussian PDF we need to allow for the 
intrinsic scatter of SN luminosities to be non-gaussian as well. Even 
if this exactly mimicked the signal at one redshift this cannot be the case at all 
redshifts, hence the data 
can distinguish between these two models due to their different redshift 
dependencies. For the PDF there are a few possible options. A popular one is 
to use gaussian mixture model, but here we will instead use a non-gaussian PDF of 
the form
\begin{equation}\label{eq:p_obs}
 P_{SNe}(x) = \mathcal{N}\left(1+{\rm erf}\left(\frac{k_3 x}{\sqrt{2}}\right)\right)\exp\left(-\frac{1}{2}|x|^{2-k_4}\right)\,.
\end{equation}
Here the parameter $k_3,k_4$ introduce a skewness and kurtosis, respectively, and $\mathcal{N}$ is a normalization constant.
Since the lensing PDF is non-gaussian (cf. Fig. Fig. \ref{fig:pdf_data} of the letter), the inclusion of extra parameters in the likelihood allows the exploration of possible degeneracies between them. We will show below that the data (in the absence of outliers discussed separately in section \ref{sec:outliers}) do not show much preference for 
the non-gaussian parameters, with a strong correlation between the two, so there is no need to explore more general 
PDFs given that even this parametrization leads to overfitting of the PDF. 

%standardization
The observed SNe distance modulus corrects the bolometric magnitude $m_B^*$ for stretch $X_1$ and color $C$
\begin{equation}\label{eq:m_obs}
m_{ob,i} = m_{B,i}^\star - (M_B - a X_{1,i} + b C_i) \,,
\end{equation}
where $a,b$ are nuisance parameters and $M_B$ is the a constant offset. In the Union 2.1 compilation the data provided has been already standardized, $a,b, M_B$ are fixed. 
The JLA compilation includes an offset correction depending on the host mass galaxy as $M_B = M_B^1$ if $M_{stellar}<10^{10}M_\odot$ and $M_B = M_B^1 + \Delta_M$ otherwise. In the JLA case, the parameter vector $\vec \theta$ also contains $a,b,\Delta_M$, which are sampled along with the other parameters. Note that $\bar m$ is degenerate with both $M_B$ and $H_0$, as $\bar D\propto 1/H_0$. Therefore we will vary $\bar m$ in $\vec\theta$, fix $M_B,H_0$ to their fiducial values and remove the mean posterior of $\bar m$, which is degenerate with both.

\begin{figure*}
 \includegraphics[width=0.999\textwidth]{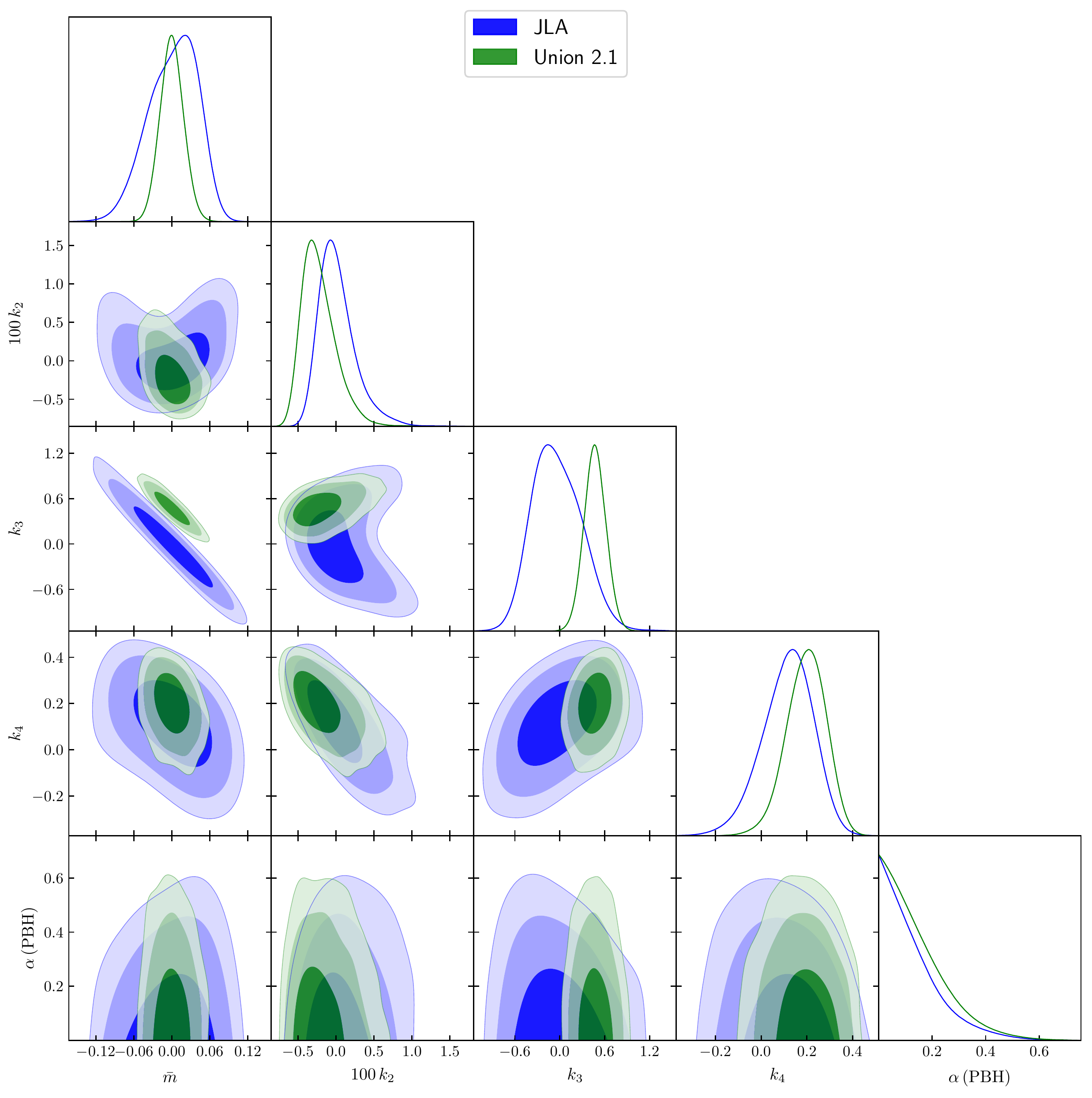}
 \caption{
 68\%, 95\% and 99\%  marginalized constraint contours for the model parameters using the JLA (blue) and Union 2.1 (green) SNe compilations. 
 Only the PBH fraction $\alpha$ and SNe population parameters in Eq. \ref{eq:p_obs} are shown.
 \label{fig:results}
}
\end{figure*}

%standard deviation
The standard deviation is corrected in quadrature for intrinsic dispersion and gravitational lensing
\begin{equation}\label{eq:eff_sigma}
\sigma^2_i = \sigma_{ob,i}^2 + k_2 - \sigma^2_{L}(z)
\end{equation}
The observational error $\sigma_{ob,i}^2$ is obtained from the diagonal of the covariance matrix, including systematics. For the JLA sample we construct the covariance matrix including the standardization parameters (equation \ref{eq:m_obs}) and their covariances, as described in Ref. \cite{Betoule:2014frx}.
SNe data includes an intrinsic magnitude dispersion to account for variability in luminosity after standarization (on top of the observational error). We correct the dispersion via $k_2$, which is allowed to have a negative sign to cancel the intrinsic scatter included in the error (and which may be model dependent).
Finally, we remove the lensing contribution $\sigma^2_{L}(z)$ to avoid double counting, as we are including realistic LSS lensing in the likelihood (equation \ref{eq:like_tot}).%

\subsection{Base Results} \label{sec:base_analysis}

\begin{table}[t!]
\vspace{10pt}
\begin{tabular} { l  c c}
 Parameter &  JLA & Union \\
\hline
{\boldmath$\bar m$       } & $0.000^{+0.068}_{-0.076}$ &  $0.000^{+0.036}_{-0.035}$ \\
{\boldmath$100\Delta\sigma^2$} & $0.02^{+0.52}_{-0.43}$ &  $-0.22^{+0.46}_{-0.39}$ \\
{\boldmath$k_3$          } & $-0.04^{+0.65}_{-0.60}$ &  $0.47^{+0.29}_{-0.29}$ \\
{\boldmath$k_4$          } & $0.12^{+0.21}_{-0.23}$ & $0.20^{+0.17}_{-0.17}$ \\
\hline
{\boldmath$a $} & $0.127^{+0.011}_{-0.011}$ & $-$ \\
{\boldmath$b $}  & $2.62^{+0.14}_{-0.14}$ & $-$ \\
{\boldmath$\Delta_M $} & $-0.047^{+0.023}_{-0.023}$ & $-$ \\
\hline 
{\boldmath$\Omega_M$     } & $0.310^{+0.011}_{-0.011}$ & $0.309^{+0.011}_{-0.012}$ \\
\hline
{\boldmath$\alpha$ (PBH)  } & $ < 0.352                  $ & $< 0.372  $ \\
\hline
\end{tabular}
\caption{ 95\% limits on the population, standarization and cosmological parameters obtained for the base analysis ($M_{\rm PBH} \gg 0.01 M_\odot$), see Fig. \ref{fig:results}. The degrading of the constraints for low PBH mass (Fig. \ref{fig:mass_dep_constraints}, letter) affects only the PBH fraction $\alpha$. \label{tab:bounds}
}
\end{table}

%cosmology depencence
\begin{figure*}[t]
 \includegraphics[width=0.48\textwidth]{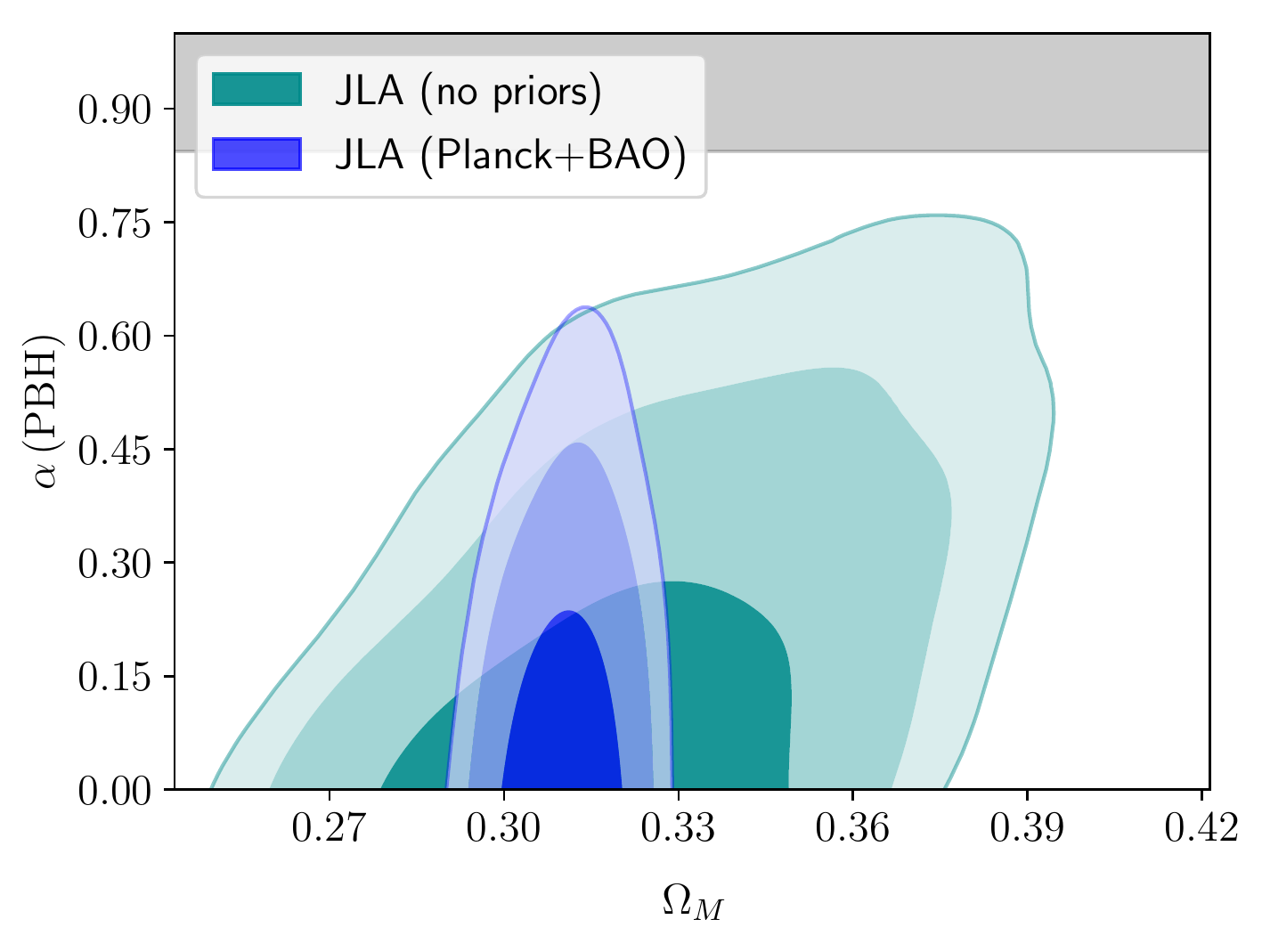}\hfill
  \includegraphics[width=0.48\textwidth]{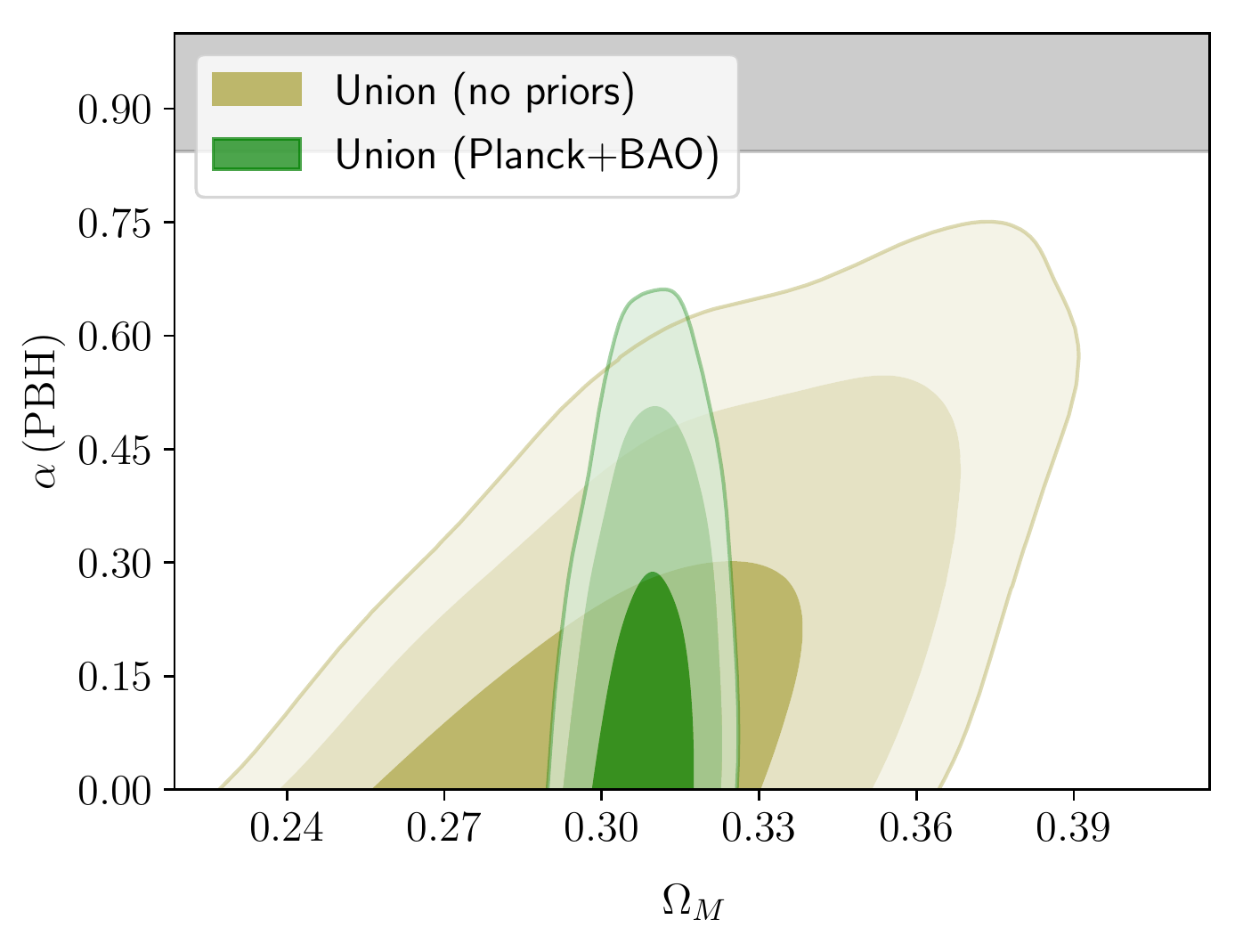}
 \caption{Effect of Planck+BAO prior on the cosmology dependence. The lighter (dashed) constraints include no prior on $\Omega_M$, removing any restriction on the flat-$\Lambda$CDM cosmology. The contours have the expected degneracy between $\Omega_M$ and the PBH fraction  $\alpha$ caused by the shift of the lensing PDF maximum associated to the empty beam.
 Contours contain 68\%, 95\% and 99\% of the probability.
 }\label{fig:cosmology_dependence}
\end{figure*}

%method
We constrain the PBH fraction by sampling the total likelihood (equation \ref{eq:like_tot}) over the parameters representing cosmology ($\alpha, \Omega_M$), the SNe population ($\bar m,k_2,k_3,k_4$), and standardization ($a,b,\Delta_M$ JLA sample only). We impose a Gaussian prior $\Omega_M = 0.309 \pm 0.006$ consistent with CMB+BAO constraints within a flat $\Lambda$CDM model \cite{Alam:2016hwk}. This prior does not use SNIa data and hence is 
uncorrelated with the PBH fraction we are constraining here. 
The likelihood (equation \ref{eq:like_tot}) was sampled using the \texttt{emcee} code \cite{emcee} and the results analyzed with GetDist \cite{getdist}. Results will be presented for the JLA \cite{Betoule:2014frx} and Union \cite{Suzuki:2011hu} datasets.

%baseline 
The bounds on the PBH fraction are $\alpha< 0.352$ (JLA) and $\alpha<0.372$ (Union)  at 95\% confidence, assuming that SNe are point sources $M_{\rm PBH}\gtrsim 1 M_\odot$. The allowed regions are presented in Fig. \ref{fig:results} and summarized on Table \ref{tab:bounds}. Both supernova samples produce consistent results, with slight variations in the parameters characterizing the SNe population: the Union sample is best fit by a non-zero skewness $k_3=0.47\pm{0.29}$, while the JLA sample exhibits a significant degeneracy between $k_3$ and the mean $\bar m$. 
%degeneracies with alpha
None of the parameters is strongly correlated with the PBH fraction $\alpha$ because 1) it produces redshift dependent effects and 2) low redshift supernovae are very effective at fixing the population parameters (using only low-$z$ SNe very similar constraints on $\bar m,k_2,k_3,k_4$ as the full analysis).
The constraints on $\Omega_M$ are entirely dominated by the external prior, and will be therefore not shown.
Our results for $\alpha$ are consistent with the forecasted sensitivities estimated in Ref. \cite{Seljak:1999tm} given our supernova sample size ($\sigma_\alpha = 20\%$ for 100 SNe at $z=1$ and $\sigma_\alpha = 5\%$ for 1000 SNe at $z=1$, assuming  $\sigma_m = 0.14$).

%outlier event rates
Our statistical analysis uses information from the entire PDF. 
The constraints on PBH fraction come both from the absence of the shift in the 
peak of PDF and from the absence of highly magnified SNe events. To get a sense of the importance of these events for the constraints, Fig. \ref{fig:pdf_data} of the letter shows that for $\alpha=1$ one expects $\sim 8$ events with $\Delta \mu >0.45$, while none are observed. 
Under Poisson distribution the probability for this is $e^{80}$ and the model is strongly constrained from the absence of these events predicted by the model. Contrary to claims in \cite{Garcia-Bellido:2017} the relevant quantity is not the number of observed overluminous events, which is 0 for $\Delta \mu >0.45$, but the number of predicted events, which is $\sim 7$ for the PBH model with $\alpha=0.84$. 
Note that in \ref{fig:pdf_data} the measurement errors are scaled to $\sigma_\mu=0.15$: the real likelihood accounts for SNe with smaller errors being more significant and those with larger errors having less weight in the final outcome.

\begin{figure*}[t]
\begin{center}
  \includegraphics[width=0.48\textwidth]{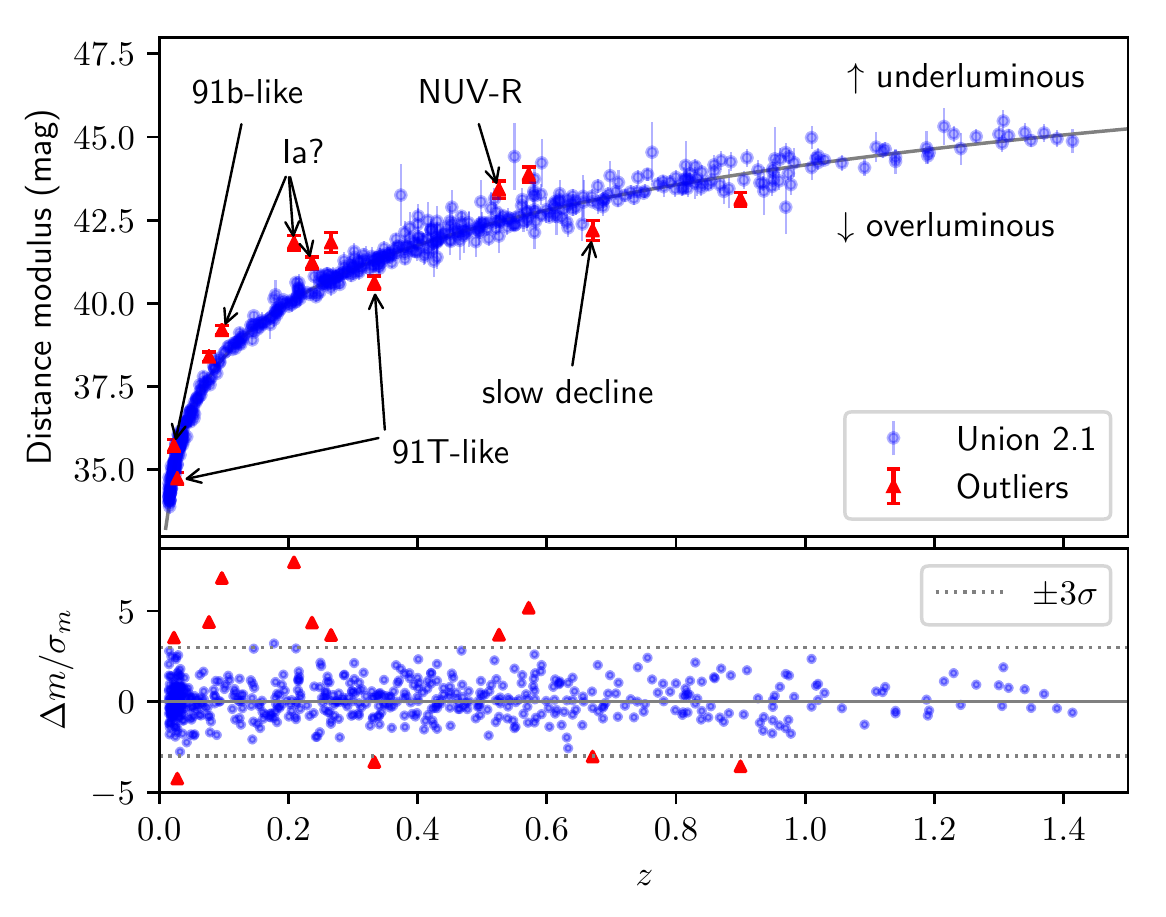} \hfill
  \includegraphics[width=0.49\textwidth]{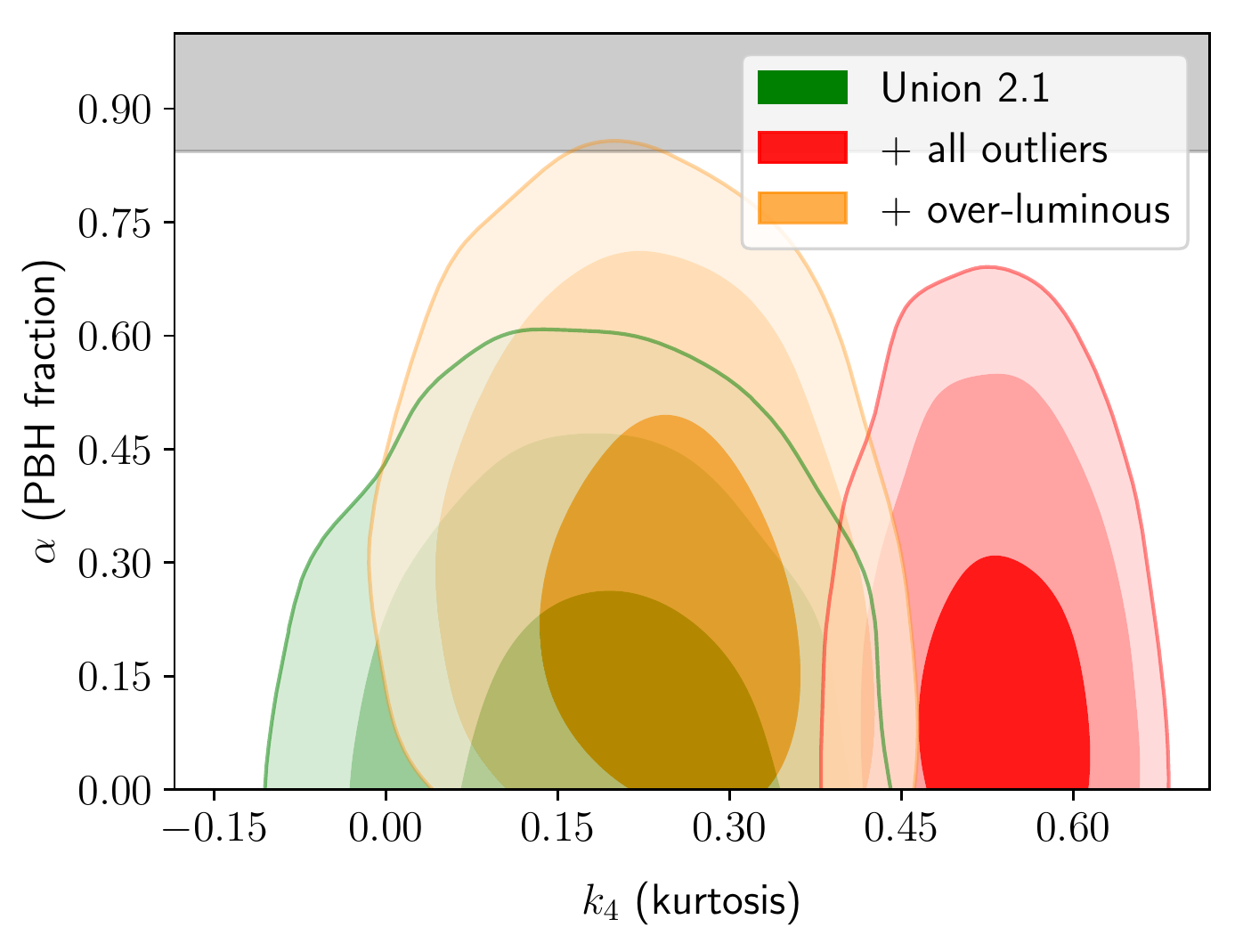}
\end{center}
 \caption{\textbf{Right panel:} Outliers in the Union 2.1 Sample. The outliers are predominantly underluminous, except at high redshift (where they would be too faint to be detectable). This is opposed to what is expected in the PBH model, which predicts only overluminous outliers. 
 \textbf{Left panel:} 68\%, 95\% and 99\%  marginalized constraint contours for the PBH fraction $\alpha$ and kurtosis $k_4$: including all outliers (red), only the overluminous ones (yellow) and none (green). Note that 3/4 of the overluminous SNe have been catalogued as peculiar type Ia and are hence unlikely to be highly-magnified events. 
 }\label{fig:outliers_union}
\end{figure*}

%size and mass distribution
The finite SNe size degrades the constraints if the PBH mass is sufficiently low. We derive constraints for monochromatic mass distributions $M_{\rm PBH}/M_\odot = 100, 10, 1, 0.1, 0.03 ,0.01, 0.003$ and $0.001$ using our effective PBH fraction prescription (equation \ref{eq:alpha_eff}). This method is validated by the lensing PDF for finite SNe, see \cite{Pei:1993}. 
The results, shown in Fig. \ref{fig:mass_dep_constraints} of the letter, indicate that the constraints degrade for $M_{\rm PBH} \lesssim 10^{-3} M_\odot$ but remain basically unaffected and consistent for higher masses. 
The constraints on the extended mass distributions 
depend on the fraction of the mass in PBH with $M<10^{-2}M_{\odot}$. 
For example, for $P(M)$ described as a 
log-normal distribution with mean $\mu=2.5M_{\odot}$ and log-normal 
variance of $\sigma=1.15$ advocated in \cite{Garcia-Bellido:2017}, we find $\bar\alpha$ increases negligibly, since 
for this model only a negligible fraction ($\sim 10^{-6}$) of PBH has $M<10^{-2}M_{\odot}$.

Lifting the Planck+BAO prior on $\Omega_M$ leads to a slight degrading of the overall constraints, with $\alpha < 0.440$ (JLA) and $\alpha < 0.437$ (Union) at 95\% c.l., see Fig. \ref{fig:cosmology_dependence}. 
The degeneracy between $\alpha,\Omega_M$ exists due to the effect of $\Omega_M$ on the luminosity distance, which is partly degenerate with the empty-beam demagnification effect.
Note that Union 2.1 leads to a slightly lower value of $\alpha$ because the central value of $\Omega_M$ prefered by the dataset is slightly lower than for JLA.
The constraints remain effective because of the different redshift dependence of both effects and the lack of highly-magnified supernovae, which is independent of the background expansion.
In the best case scenario for PBH the degeneracy allows for $\alpha\lesssim 0.60$ (95 \% c.l.) for $\Omega_M\sim 0.36$, which excluded at about $8$ standard deviations by Planck+BAO.

\section{Systematic effects} \label{sec:systematics}

\subsection{SNe Outliers}\label{sec:outliers}

% SNe outliers are expected to contaminate the sample. 
Non-standardizable SNe are expected to contaminate the type Ia sample, and several classes of peculiar Type Ia supernovae have been identified. These classes include very bright (overluminous) objects (e.g. 91T-like \cite{1992ApJ...384L..15F,1992AJ....103.1632P}, super-Chandrasekhar \cite{Howell:2006vn,Scalzo:2012yp} and Ia-CSM \cite{Hamuy:2003rh,Dilday:2012cy,Silverman:2013zsa}) as well as intrinsically fainter (underluminous) ones (e.g. 91bg-like), see Ref. \cite{Taubenberger:2017hoo} for a review.
A critical prediction of the PBH model on SNe lensing is the existence of highly magnified events. If existing, these events would be removed from the sample by the outlier rejection procedure and hence bias the result against the PBH model. 
The simplest approach to outliers is simply ``clipping'' data points whose residual over the prediction is above some threshold, i.e. those points that deviate beyond what is expected by lensing magnification (cf. section 2.1 of \cite{Conley:2011ku}). Other prescriptions are more sophisticated, including the fitting to two mixed SNe populations, one consisting of ``real'' type Ia events, and an outlier population with a large scatter \cite{Kunz:2006ik}.
Note that other sample-selection methods (i.e. cuts based on light-curves or spectroscopy) are generally not affected by gravitational lensing (an exception is microlensing-induced time dependence \cite{Foxley-Marrable:2018dzu}, but that is only important for light PBH masses, cf. Fig. 1 of \cite{Metcalf:2006ms}).

In order to test the effect of outliers we consider the supernovae rejected in compiling the Union 2.1 dataset (see also Ref. \cite{Rubin:2015rza}), see Fig. \ref{fig:outliers_union}, where a visual inspection shows that their distribution is not compatible with the expectations of the PBH model. A detailed review of the SNe sample shows that most supernovae have features that identify them as outliers: three overluminous (91T-like, or very slow decaying SNe $z=0.02, 0.33, 0.67$ \cite{Foley:2008qc}), and five underluminous (91bg-like $z=0.02$ \cite{Ganeshalingam:2012yj},
uncertain ``Ia?'' classification $z=0.10,0.21,0.24$ \cite{Zheng:2008fn} or NUV-red $z=0.53$). This would leave only one overluminous ($z=0.9$) and three underluminous ($z=0.08,0.27,0.57$) SNe that can not be discarded based on additional information.%
\footnote{We thank David Rubin for providing the Union 2.1 outliers and Lluis Galbany for a detailed assesment of the outliers.}

\begin{figure*}[t!]
\includegraphics[width=0.495\textwidth]{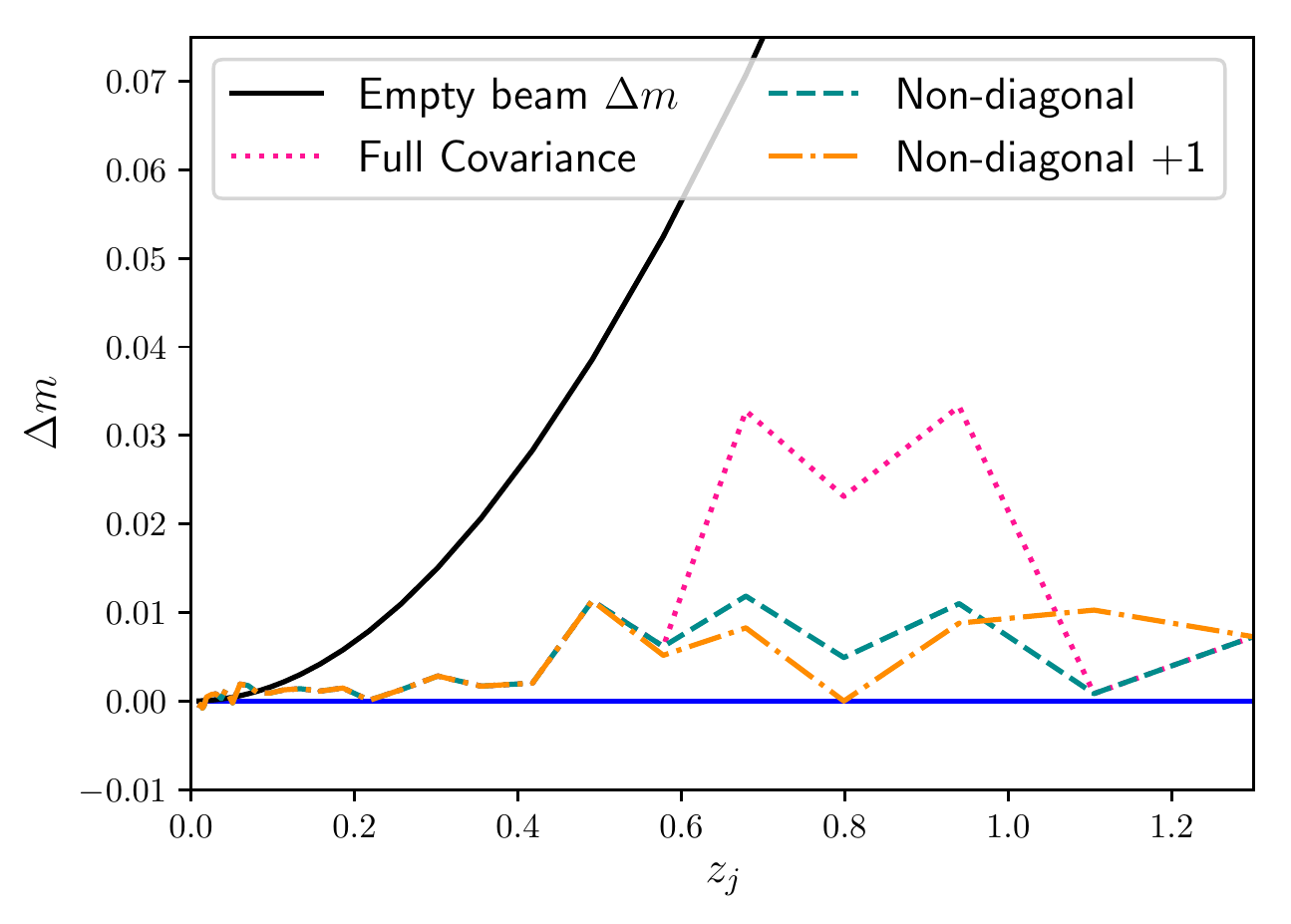}
 \includegraphics[width=0.48\textwidth]{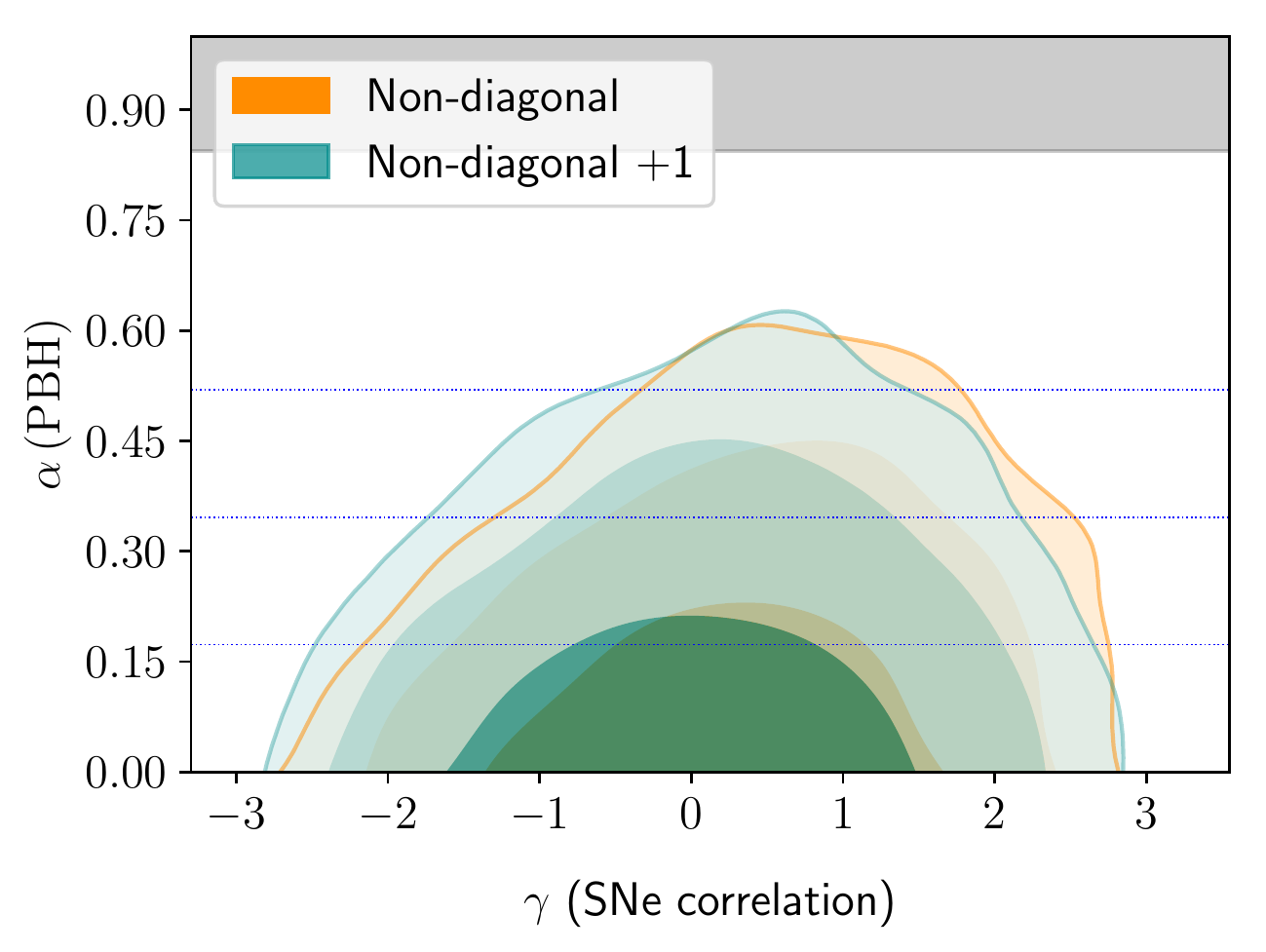}
 \caption{Effects of correlated noise on the PBH constraints following the model in Sec. \ref{sec:sne_covariance}. \textbf{Left panel:} correlated noise model (equation \ref{eq:correlated_noise}) with different assumptons on the compressed JLA likelihood. The empty beam demagnification (translated to magnitudes) is shown for comparison.
\textbf{Right panel:} 68\%, 95\% and 99\%  marginalized constraint contours for the PBH abundance and the nuisance correlation parameter $\gamma$ (Eq. \ref{eq:correlated_noise}). Using the correlation matrix without the diagonal terms (orange) weakens the constraints more than the case without the diagonal and next-to-diagonal (dark cyan). The horizontal blue lines mark the 1,2 and 3 $\sigma$ constraints in the standard analysis ($\gamma=0$).
 \label{fig:correlated_noise}
}
\end{figure*}

The predictions in Figs. \ref{fig:pbh_signatures} and \ref{fig:pdf_data} of the letter show that the PBH model predicts overluminous events, without any underluminous counterpart.
In contrast, Union 2.1 outliers are predominantly underluminous, and have an asymmetric distribution: underluminous SNe deviate as much as $\Delta m/\sigma_m \sim 8$, while overluminous SNe only depart as much as $\Delta m/\sigma_m \sim -4 $. 
We note that this is unlikely due to selection effects since overluminous outliers would be much easier to detect. Finally, the four overluminous outliers 
are uniformly distributed over the entire redshift range, while PBH scenario would prefer these events to be predominantly at higher redshifts. 

%outliers
Including all outliers in the Union sample modifies the constraints to $\alpha < 0.440$ (95\%), only a $15\%$ difference. This is because the data are not compatible with the peak of the lensing PDF being at the empty beam distance (equation \ref{eq:empty_beam}), nor compatible with the predicted fraction of magnified events, see Fig. Fig. \ref{fig:pdf_data} of the letter. In addition, the majority of outliers are underluminous, with only a few overluminous cases, cf. Fig. \ref{fig:outliers_union}.
The main difference in the analysis with outliers is the preference towards a non-zero kurtosis $k_4 = 0.537^{+0.094}_{-0.10}$, different from zero at $\approx 5\sigma$ significance, see Fig. \ref{fig:results}.
The interplay between the outliers and higher order SNe population parameters highlights the importance of including a sufficiently detailed model, e.g. equation (\ref{eq:p_obs}).
For the sake of completeness, we performed the analysis including only the overluminous outliers, even though 75\% of those SNe are likely to be peculiar SNe based on either spectroscopic or lightcurve features. Ignoring this information and interpreting them as magnified events (while rejecting all the underluminous outliers) leads to a bound on the PBH fraction to $\alpha < 0.581$ (95\%), which is still a significant rejection of PBH accounting for the totality of DM.

\subsection{Correlated noise} \label{sec:sne_covariance}

Correlations in the noise model arise due to the correlations in the 
calibration errors, but are difficult to include in the non-gaussian likelihood (equation \ref{eq:like_tot}). 
We will model correlations by adopting a rank one (plus diagonal) covariance matrix approximation. We
model the observed magnitude in equation (\ref{eq:m_obs}) by adding to it a redshift dependent term
that has been extracted from the rank one decomposition of the 
covariance matrix
\begin{equation}\label{eq:correlated_noise}
m_{c} = \gamma \cdot e(z)\,, \quad \text{with} \quad
 e(z_i) = \frac{1}{N_J}\sum_{j\in J} \frac{\bar C_{ij}}{\sqrt{C_{jj}}}\,. 
\end{equation}
Here $\bar C_{ij}$ is the simplified JLA covariance matrix (Table F2 in Ref. \cite{Betoule:2014frx}) but setting the diagonal and next-to-diagonal elements to zero ($\bar C_{ii} = \bar C_{i, i\pm 1} =0$) as they represent the combined error at each redshift and the correlations induced by the data compression. 
For the sake of comparison we will also discuss the effect of using the diagonal-free ($\bar C_{i, i\pm 1} \neq 0$) and the full covariance matrix ($\bar C_{i i} \neq 0$).
The sum is over $J=(27,28,29)$, and normalized by $N_J=3$: this gives an average over the redshifts $z = (0.679,0.799,0.940)$ on which the sensitivity to the PBH fraction is strongest, while discarding the last entries to avoid auto-correlations (accounted in the measurement errors) and next-neighbor correlations (induced by data compression). The redshift dependence of these models is shown in Fig. \ref{fig:correlated_noise}.
In the model  of equation (\ref{eq:correlated_noise}) $\gamma$ is a free parameter to be sampled with a Gaussian prior of width $\sigma_\gamma = 1$.

%Results
%correlations
The results are robust against the inclusion of correlated noise, which weakens the constraints only slightly. The correlation model based on the JLA simplified covariance matrix (equation \ref{eq:correlated_noise}) changes the 95\% confidence bounds to $\alpha< 0.363$ when both the diagonal and elements next to the diagonal are removed and $\alpha < 0.372$ when only the the diagonal is removed, see Fig. \ref{fig:correlated_noise}.
Using the full correlation matrix (including the diagonal, which artificially increases the errors) degrades the bounds to only $\alpha < 0.398$. None of these different prescriptions are sufficient to significantly weaken the constraints on the fraction of primordial black holes.

\subsection{Selection bias and SNe evolution} \label{sec:selection_bias}

Another potential issue is the selection bias in PBH models: high-redshift supernovae may be selectively brighter. 
This selection (or Malmquist) bias is usually corrected through a frequentist procedure where the mean 
bias effect is simulated and corrected for, separately for each of the surveys (e.g. SDSS, SNLS...). The overall effect can be difficult to visualize, since 
the color and stretch parameters entering the standardization relation (\ref{eq:m_obs}) are correlated with the intrinsic magnitude, all of 
which can play a role in selection bias effects. Despite the 
large size of the selection bias effect on color at the upper redshift range of any specific survey
(Fig. 11 of \cite{Betoule:2014frx}), the overall bias correction is small
(less than 0.04 in $\Delta m$, Fig. 5 of \cite{Betoule:2014frx}).

It is important to note that the different surveys give consistent results 
with each other after correction (Fig. 11 of \cite{Betoule:2014frx})
despite the fact that they can have a large color selection bias. 
This gives confidence 
that the selection effects are properly corrected over the range of models we are considering here. 
We note that the variation in peak value of $\Delta m$ we are considering here are up to 0.1, which is a smaller change than 
the differences between the accelerating and non-accelerating universe models, so as long as the bias 
correction is valid for the standard cosmological models it should also be 
valid for our lensing models.

Our analysis assumed negligible evolution of the SNe population (Eq. \ref{eq:p_obs}). 
While SNe evolution (e.g. $z$-dependent $\bar m, k_2,k_3,k_4$) is likely to weaken our bounds, it is unlikely to modify our conclusions significantly. 
For instance, a redshift-dependent mean $\bar m$ is degenerate with the background cosmology. 
This case is very similar to the analysis without priors on $\Omega_m$ (Fig. \ref{fig:cosmology_dependence}), where lifting the prior weakens the bounds but the model remains constrained due to the lack of magnified SNe.

SNe population evolution can not compensate the lack of magnified events because  magnification is uncorrelated with the SNIa properties (e.g. the form of Eq. \ref{eq:like_tot}).
This lack of correlation is due to intrinsic SNe luminosity being a local property, while on the other hand gravitational magnification depends on the cosmological matter distribution along the line of sight, which is completely independent from the source.
Therefore, the absence of magnified SNe can not be explained by a more general, redshift dependent intrinsic luminosity distribution generalizing Eq. \ref{eq:p_obs}.

\bibliography{bhsne_references}

\end{document}